\documentclass[%
 aip,
 amsmath,amssymb,
 reprint,%
nofootinbib,
]{revtex4-1}

\usepackage{graphicx}% Include figure files
\usepackage{dcolumn}% Align table columns on decimal point
\usepackage{bm}% bold math
\usepackage[utf8]{inputenc}
\usepackage[T1]{fontenc}
\usepackage{mathptmx}
\usepackage{etoolbox}
\usepackage{graphicx}% Include figure files
\usepackage{dcolumn}% Align table columns on decimal point
\usepackage{bm}% bold math
\usepackage{hyperref}% add hypertext capabilities
\usepackage{physics}
\usepackage{siunitx}
\usepackage[dvipsnames]{xcolor}
\usepackage{empheq}
\usepackage{subcaption}
\usepackage{soul}
\usepackage{enumerate}

\makeatletter
\def\@email#1#2{%
 \endgroup
 \patchcmd{\titleblock@produce}
  {\frontmatter@RRAPformat}
  {\frontmatter@RRAPformat{\produce@RRAP{*#1\href{mailto:#2}{#2}}}\frontmatter@RRAPformat}
  {}{}
}%
\makeatother
\newcommand{\iu}{{i\mkern1mu}}

\begin{document}

\preprint{AIP/123-QED}

\title{Exact Solution for a Real Polaritonic System Under Vibrational Strong Coupling in Thermodynamic Equilibrium: Absence of Zero Temperature and Loss of Light-Matter Entanglement
}
% Force line breaks with \\
\author{Dominik Sidler}
  \email{dsidler@mpsd.mpg.de}
 \affiliation{Max Planck Institute for the Structure and Dynamics of Matter and Center for Free-Electron Laser Science, Luruper Chaussee 149, 22761 Hamburg, Germany}
\affiliation{The Hamburg Center for Ultrafast Imaging, Luruper Chaussee 149, 22761 Hamburg, Germany}%Lines break automatically or can be forced with \\

\author{Michael Ruggenthaler}
  \email{michael.ruggenthaler@mpsd.mpg.de}
  \affiliation{Max Planck Institute for the Structure and Dynamics of Matter and Center for Free-Electron Laser Science, Luruper Chaussee 149, 22761 Hamburg, Germany}
  \affiliation{The Hamburg Center for Ultrafast Imaging, Luruper Chaussee 149, 22761 Hamburg, Germany}

\author{Angel Rubio}
  \email{angel.rubio@mpsd.mpg.de}
  \affiliation{Max Planck Institute for the Structure and Dynamics of Matter and Center for Free-Electron Laser Science, Luruper Chaussee 149, 22761 Hamburg, Germany}
    \affiliation{The Hamburg Center for Ultrafast Imaging, Luruper Chaussee 149, 22761 Hamburg, Germany}
  \affiliation{Center for Computational Quantum Physics, Flatiron Institute, 162 5th Avenue, New York, NY 10010, USA}
  \affiliation{Nano-Bio Spectroscopy Group, University of the Basque Country (UPV/EHU), 20018 San Sebasti\'an, Spain}

\date{\today}% It is always \today, today,

\begin{abstract}
%When light and matter strongly couple, both subsystems are modified. Strong light-matter coupling is therefore a very appealing approach to change and control properties of light and matter alike and to engineer quantum entanglement. Commonly, however, only one of these subsystems is treated in detail, while the other is usually treated approximately. In this work we consider light and matter on equal footing and determine the numerically exact quantum thermal state of a chemically realistic molecule (HD+) strongly coupled to an optical cavity. We discuss how strong coupling could potentially provide access to hybrid quantum statistics, show the emergence of non-canonical behavior of the different subsystems and highlight how light-matter quantum entanglement is quickly lost when the temperature is increased. We use these specific results to discuss under which conditions one might be able to employ ro-vibrational states of molecules, which are commonly only considered as a source for decoherence, for quantum technological applications. While distillable entanglement is lost for higher temperatures, increased quantum thermal fluctuations might still be observed even at ambient conditions. This has interesting implications for recent experimental results within polaritonic chemistry, where chemical reactions can be modified even at room temperature. Indeed, we believe that the increased quantum thermal fluctuations in vibrational strong coupling can be used to generate non-classical thermal ensembles at ambient conditions.
The first exact quantum simulation of a real molecular system (HD$^+$) under strong ro-vibrational coupling to a quantized optical cavity mode in thermal equilibrium is presented. Theoretical challenges in describing strongly coupled systems of mixed quantum statistics (Bosons and Fermions) are discussed and circumvented by the specific choice of our molecular system. Our exact simulations reveal the absence of a zero temperature for the strongly coupled matter and light subsystems, due to cavity induced non-equilibrium conditions. Furthermore, we explore the temperature dependency of light-matter quantum entanglement, which emerges for the groundstate, but is quickly lost already in the deep cryogenic regime, opposing predictions from phenomenological models (Jaynes-Cummings). Distillable molecular light-matter entanglement of ro-vibrational states may open interesting perspectives for quantum technological applications. Moreover,  we find that the dynamics (fluctuations) of matter remains modified by the quantum nature of the thermal and vacuum field fluctuations for significant temperatures, e.g. at ambient conditions. These observations (loss of entanglement and coupling to quantum fluctuations) has far reaching consequences for the understanding and control of polaritonic chemistry and materials science%in polaritonic chemistry
, since a semi-classical theoretical description of light-matter interaction becomes feasible, but the typical canonical equilibrium assumption for the nuclear dynamics remains broken. This opens the door for quantum fluctuations induced stochastic resonance phenomena %in polaritonic systems 
under vibrational strong coupling. A plausible theoretical mechanism to explain the experimentally observed resonance phenomena in absence of periodic driving, which have not yet been understood.%\revDS{Ruggi: do you have some abstract modifications in mind to focus less on chemistry, but also on materials and quantum information?}
\end{abstract}

\maketitle

Strong coupling of quantum light and matter via optical cavities has become a rapidly developing technique, which has made an outstanding impact across scientific disciplines over the last years. For example, exciton-polaritonic condensates have attractive features for quantum computing,\cite{ghosh2020quantum} or cavity magnon polariton systems are promising candidates for quantum information processing with long spin coherence times.\cite{rao2019analogue} Furthermore, modifications of the transition temperature of superconductors were predicted\cite{sentef2018cavity,curtis2019cavity,schlawin2019cavity} and measured,\cite{thomas2019exploring} novel optical devices for wavefront engineering and sub-wavelength focusing became feasible\cite{chaudhary2019polariton}. Furthermore, the cavity induced stabilisation of the ferroelectric phase in SrTiO$_3$ or the magnetic control of proximate spin liquid $\alpha$-RuCl$_3$ have been proposed.\cite{latini2021ferroelectric,bostroem2022} Large scientific attention was also created in the chemistry community, due to successful inhibition,\cite{Thomas2016} steering\cite{thomas2019tilting} and enhancing\cite{lather2020improving} molecular reaction rates under strong vibrational coupling condition.
 
The decisive ingredient of these experiments is that matter couples strongly to the vacuum or few thermally populated photons of a cavity instead of weakly coupling to many photons under external laser driving. In the latter case only transient (Floquet-type) non-equilibrium states can emerge, which are hard to detect experimentally due to decoherence, dissipation and heating effects~\cite{lloyd20212021}. For the strongly-coupled cavity-matter system, however, robust thermal equilibrium states of light and matter emerge, which are of significant importance for the physics under investigation (polaritonic states and polaritonic quantum matter).\cite{basov2021polariton} 
The theoretical description of quantized light and matter under strong coupling conditions is a notoriously hard problem to tackle, as it \textit{a priori} requires a quantum electrodynamics (QED) description in full thermal equilibrium.
To bypass this complexity, phenomenological  models are used predominantly, instead of an ab initio QED description. These models have been devised in quantum optics (e.g. Jaynes-Cummings) and are designed to model photon properties accurately,\cite{jaynes1963comparison} but at the same time impose strong simplifications for the involved matter subsystem, i.e. the detailed properties of the matter subsystem are assumed irrelevant except for their influence on the light field. Only recently the reverse question, i.e., how the strongly-coupled photons influence matter properties, has become the focus of intensive research in polaritonic or QED chemistry and materials science~\cite{ebbesen2016,flick2017atoms,kockum2019ultrastrong,Hubener2020}. However,   the details of the photon field and an accurate description of the coupled thermal equilibrium are commonly assumed to be irrelevant in this matter-driven perspective and thus $T=0$ is commonly assumed. Yet all of which specialized viewpoints seem to be insufficient to explain certain experimental observations, such as the resonance condition for suppressing chemical reactions via strong coupling~\cite{Thomas2016, sidler2021perspective} or how strong coupling can influence complex aggregation processes of molecule-metal complexes~\cite{hirai2021selective}. Much experimental effort has thus been spent into engineering the matter subsystem to enforce the (unfortunate) theoretical quantum-optical simplifications, e.g., that molecules become effective two-level systems~\cite{PhysRevLett.127.133603,toninelli2021single}.

Here we try to unify these specialized viewpoints based on rigorous theoretical ground, i.e. based on the stationary solution of the exact Quantum-Liouville equation in the non-relativistic Pauli-Fierz limit of QED. We then deduce fundamental properties from a paradigmatic molecular test system (HD$^+$) considering the full (chemical) complexity, i.e., by having light and matter treated fully quantized and including also the coupling to an external heat bath. In particular we will address the following questions: How does the temperature of the total ensemble translate to the individual subsystems? A common simplification is to assume that the effective temperature of the subsystems is equivalent to the temperature of the total ensemble. How are the quantum and thermal fluctuations of the subsystems related? Again, a common simplification is to assume that the fluctuations of the subsystems stay unaffected and can be replaced by the fluctuations of the uncoupled systems. Finally, is light and matter quantum-entangled and what happens to the entanglement when we increase the temperature? While it is commonly accepted that quantum entanglement should be lost with increasing temperature, a detailed quantification for realistic systems is usually not available. Indeed,  the viewpoint of  collective "super-molecules"\cite{galego2015cavity,galego2017many}  (formed by light and matter at ambient conditions) seems to be contradictory, which is a widely spread concept within polaritonic community. In this context also the question of how to define the thermal state and quantum-statistics of a collectively-coupled ensemble of molecules will become important. 
Particularly interesting is that we will not focus on the electronic energy range, for which the common quantum-optical models have been designed, but investigate the  low-energy ro-vibrational regime instead, which is predominantly affected by temperature. Usually the ro-vibrational degrees of freedom are only considered as decoherence channels for electronic excitations and their detailed quantum-mechanical nature is not investigated for potential quantum-technological applications. Indeed, molecular systems in principle allow to go beyond simple qubit representations\cite{wernsdorfer2019synthetic,wasielewski2020exploiting} where decoherence sources can be mitigated/controlled by the specific molecular composition.\cite{krzyaniak2015fast,yu2016long,wasielewski2020exploiting,pscherer2021single,gurlek2021engineering,zirkelbach2022high}  Based on the our results we will highlight that  strongly-coupled molecule-cavity systems can show robust distillable quantum entanglement for cryogenic situations and hence such systems provide a potential platform for the development and implementation of future quantum technologies. Furthermore, for higher temperatures, where entanglement is quickly lost, non-trivial feedback between light and matter unravels  cavity induced non-equilibrium mechanisms, which become decisive in the context of polaritonic chemistry and materials science. Finally,  we will be able to develop a generic theoretical framework to describe molecular systems strongly dressed by a cavity, based on our simulation results for HD$^+$ in thermal equilibrium.

This work is structured as follows: We first discuss how we theoretically describe the quantized light-matter system in the long-wavelength limit of non-relativistic QED and show the necessary transformations to make the problem numerically tractable. Furthermore, theoretical issues for strongly coupled systems of mixed particle statistics in thermal equilibrium are addressed. In a second step, exact thermal equilibrium solutions are presented with a focus on strong-coupling-induced temperature modifications, quantum thermal fluctuations of light and matter as well as (loss of) light-matter entanglement for real molecular systems. The entanglement predictions are then contrasted to predictions from the ubiquitous Jaynes-Cummings model of quantum optics. In a third step, a concise picture of cavity induced (non)-equilibrium effects is developed and important implications for cryogenic applications are derived (e.g. quantum computing, superconductivity), as well as under ambient conditions (material science and polaritonic chemistry). We end this work with with a forward look and perspective section.

%reference JC ion laser cooling paper of Cirac et al. in Ref. \citenum{cirac1993cooling}. Notice there is lindblad term, discreticed com motion in harmonic trap??? or just from the gauge? detuning relevant, but I think the crucial difference is that correlations between COM and rest not included (doublecheck) We don't have harmonic trap in our simulations, but this had a huge impact! should we add that?
%Overall, I suspect harmonic trap can change COM temperature, but does it also change relative motion?
%Overall story could be: harmonic trap cools COM motion, whihc gets more and more inefficient the lower the temperature is, this is not true for relative motion? Is this really true! 
%Harmonic trap change temperature integral! And I think harmonic trap influences also relative coordinates!!!

\section{Exact Quantum Canonical Equilibrium Solution for HD$^+$ Molecule in a Cavity}

\subsection{Hamiltonian Representation}
In the following we rely on the non-relativistic QED Pauli-Fierz (PF) Hamiltonian in dipole approximation for the fundamental description of the light-matter interaction within a cavity tuned to the infrared or optical regime.~\cite{ryder1996quantum,craig1998molecular,spohn2004,spohn2004,ruggenthaler2014quantum,flick2015kohn,ruggenthaler2018quantum} 
%Besides the usual quantized light-matter coupling.  
The resulting Hamiltonian assumes the following form in the Coulomb gauge, \cite{ruggenthaler2015ground}
%
%For the fundamental description of quantized light-matter interaction in a  cavity we rely on the Pauli-Fierz Hamiltonian in the coulomb gauge can be written as in the Coulomb gauge,\citenum{ruggenthaler2015ground} 
\begin{eqnarray}
\hat{H}&=&
\sum_{i=1}^{N_p}\frac{1}{2m_i}\big(\hat{\bold{p}}_i-Z_i\hat{\bold{A}}\big)^2 +\sum_{i<j}^{N_p}\frac{Z_i Z_j}{|\hat{\bold{r}}_{i}-\hat{\bold{r}}_{j}|}+\sum_{\alpha}\omega_{\alpha}\bigg(\hat{a}_{\alpha}^\dagger\hat{a}_{\alpha}+\frac{1}{2}\bigg),
\label{eq:PFHeigfull}
\end{eqnarray}
where $N_p$ is the number of (fermionic or bosonic) massive particles, i.e., electrons and effective nuclei that constitute the molecules inside the cavity, with $m_i$ and $Z_i$ the corresponding masses and charges, respectively. For each particle we denote the conjugate self-adjoint momentum and position operators as $\hat{\bold{p}}_i$ and $\hat{\bold{r}}_i$, respectively. The photonic environment is defined in terms of modes $\alpha$ with corresponding frequency $\omega_{\alpha}$, linear polarization direction $\boldsymbol{\epsilon}_{\alpha}$ and coupling strength (effective mode volume) $\lambda_{\alpha}$. Here $\hat{a}_{\alpha}^{\dagger}$ is the usual bosonic creation and $\hat{a}_{\alpha}$ the annihilation operator for mode $\alpha$. The quantized transverse vector potential is then given as
%with polarisation direction $\lambda$, $\omega_\bold{n}=|\bold{k}_\bold{n}|=|2\pi \bold{n}/L|$ with mode volume $L$ and transversal polarization vectors $\epsilon_{n,\lambda}$.  The usual canonical position and momentum operators of particle $i$ (electron or nuclei) with mass $m_i$ and charge $Z_i$ are given by $\hat{\bold{r}}_i, \hat{\bold{p}}_i$. The number of involved fermionic matter particles is denoted by $N_p$
%The quantized vector potential operator reads explicitly \cite{},
\begin{eqnarray}
\hat{\bold{A}}=\sum_{\alpha} \frac{\lambda _{\alpha}\boldsymbol{\epsilon}_{\alpha}}{\sqrt{2\omega_{\alpha}}}(\hat{a}_{\alpha} +\hat{a}_{\alpha}^\dagger),
\end{eqnarray}
%with vacuum dielectric constant $\epsilon_0$ and the usual creation $\hat{a}^\dagger$ and annihilation $\hat{a}$ operators for the photons acting on the bosonic Fock space.
We have  neglected explicitly spin-dependent terms such as Zeeman and Spin-Orbit coupling terms here. The spins of the massive particles become important only for determining the symmetry of the eigenfuntions, i.e., fermionic anti-symmetry and bosonic symmetry under exchange of spin-space coordinates.  %Spin-degrees of freedoms are not included, as we will deal with distinguishable particles only in the following and we neglect any contributions arising from the Zeeman or Spin-Orbit coupling.
However, in order to make the eigenvalue problem posed by Eq.~\eqref{eq:PFHeigfull} numerically tractable, a few more simplifications are necessary: First, we restrict to one effective mode $\alpha$ of the cavity.  
%such that we have
%\begin{eqnarray}
%\hat{H}&=&
%\sum_{i=1}^{N_p}\frac{1}{2m_i}\big(\hat{\bold{p}}_i-Z_i\frac{\boldsymbol{\lambda}}{\sqrt{2 \omega}}(\hat{a} + \hat{a}^{\dagger})\big)^2 +\sum_{i<j}^{N_p}\frac{Z_i Z_j}{|\hat{\bold{r}}_{i}-\hat{\bold{r}}_{j}|}\nonumber\\
%&&+\sum_{\alpha^\prime}\omega_{\alpha^\prime}\bigg(\hat{n}_{pt,\alpha^{'}}+\frac{1}{2}\bigg),
%\label{eq:PFHeigfull}
%\end{eqnarray}
%by restricting the coupling to a linearly polarised cavity along $\bold{\lambda}_\alpha$, with one effectively coupled mode $\alpha$. The photon number operator is defined as $\hat{n}_{pt,\alpha}=\hat{a}_{\alpha}^\dagger \hat{a}_{\alpha}$ with corresponding mode frequency $\omega_\alpha$. %and creation / annihilation operators $\hat{a}^\dagger,\hat{a}$. 
%The canonical photon displacement field operators can be written as $\hat{q}_\alpha:=\sqrt{\frac{1}{2\omega_\alpha}}(\hat{a}_{\alpha}^\dagger+\hat{a}_{\alpha})$ and $\hat{p}_\alpha:=\iu\sqrt{\frac{\omega_\alpha}{2}}(\hat{a}_{\alpha}^\dagger-\hat{a}_{\alpha})$. Notice that the photon modes $\alpha^\prime $ effectively decouple from the system and will be neglected subsequently. However, they become relevant when discussing emerging modifications of the black body radiation due to the cavity (see e.g. Ref. \citenum{?}) %However, we keep them for the moment to relate our equilibrium results to the black body radiation later on. ADJUST
As a next step we restrict to three particles, i.e., $N_p=3$. This allows us to treat, e.g., a helium atom, an H$_2^+$ or an HD$^+$ molecule~\cite{sidler2020chemistry}. Here we choose an HD$^+$ molecule, that is, a positively charged molecule with one proton, one deuteron and one electron. 
%Despite the simplifications made so far, the stationary eigenvalue problem imposed by the Hamiltonian in  Eq. (\ref{eq:PFHeigfull}) typically remains computationally inaccessible for real molecular setups. First, numerically exact results were only recently presented for $N_p=3$ constituents (e.g. He, HD+ or H$_2^+$). 
Going beyond three particles is numerically  intractable with nowadays computational power except for the introduction of additional approximations, as for example done by exchange correlation functionals in QEDFT~\cite{ruggenthaler2014quantum,pellegrini2015optimized,flick2016exact,schafer2021making,flick2021simple} or by QED coupled cluster methods~\cite{mordovina2019polaritonic,haugland2020coupled,pavosevic2021polaritonic}. Having numerically exact eigenvalues and eigenstates available for HD$^+$ will subsequently allow us to investigate exact thermodynamic equilibrium properties and light-matter entanglement under ro-vibrational strong coupling. For this purpose, we shortly recapitulate the key technical ingredients of our problem-adapted numerical approach, as they become essential for the subsequent discussions.

To achieve a numerically tractable form of our quantized 3-body problem coupled to one quantized cavity-photon mode in the long-wavelenth limit, the corresponding non-relativistic Pauli-Fierz Hamiltonian has to be expressed in centre-of-mass (COM) $\bold{R}_c:= \tfrac{\sum_i m_i \bold{r}_i}{\sum_i m_i}$ and relative coordinates $\bold{r}_{ci}= \bold{r}_{i} - \bold{R}_c$. Moreover, a relative velocity form of the Hamiltonian becomes important,~\cite{sidler2020chemistry} which is obtained from a unitary Power-Zienau-Woolley transformation%~\cite{pzw}
\begin{eqnarray}
\hat{U}:=\exp\left(\iu \frac{\lambda_{\alpha} \boldsymbol{\epsilon_\alpha}\cdot\hat{\boldsymbol{d}}}{\sqrt{2 \omega_{\alpha}}}(\hat{a}_{\alpha} + \hat{a}^{\dagger}_{\alpha})\right),\label{eq:pzw}
\end{eqnarray}
where the relative dipole operator was introduced as,
\begin{eqnarray}
\hat{\boldsymbol{d}}:=\sum_{i=1}^3 Z_i \bold{\hat{r}}_{ci}.
\end{eqnarray}
Next we perform a canonical commutator-preserving substitution $S$ of the photon operators, i.e., $\hat{a}_{\alpha}\overset{S}\mapsto -\iu \hat{a}_{\alpha}$ and $\hat{a}_{\alpha}^\dagger\overset{S}\mapsto\iu \hat{a}^\dagger_{\alpha}$, resulting in~\cite{sidler2020chemistry}
\begin{eqnarray}
\hat{H}^\prime&:=&S\circ \hat{U} H\hat{U}^\dagger\\
&=&
\frac{1}{2M}\bigg(\hat{\bold{P}}_c-\frac{\lambda_{\alpha}\boldsymbol{\epsilon}_{\alpha} Q_{\mathrm{tot}}}{\omega_{\alpha}} \hat{p}_{\alpha}\bigg)^2 +\sum_{i=1}^3 \frac{\hat{\bold{p}}_{ci}^2}{2m_i}+\sum_{i<j}^{3}\frac{Z_i Z_j}{|\hat{\bold{r}}_{ci}-\hat{\bold{r}}_{cj}|} \label{eq:HDplusHamiltonian}\\
&&+\frac{1}{2}\bigg[\hat{p}_{\alpha}^{ 2}+\omega_{\alpha}^{2}\Big(\hat{q}_\alpha-\frac{\lambda_{\alpha}\boldsymbol{\epsilon}_{\alpha}}{\omega_{\alpha}}\cdot \hat{\boldsymbol{d}}\Big)^2\bigg].\nonumber
%+\sum_{\alpha^\prime\neq\alpha}^M\bigg[\frac{1}{2}\big(\hat{p}_{\alpha^\prime}^{ 2}+\omega_{\alpha^\prime}^{2}\hat{q}_{\alpha^\prime}^2\big)\bigg]
\label{eq:PFHeiglenght}
\end{eqnarray}
%In the following we neglect the uncoupled photon modes $\alpha^\prime\neq \alpha$ and focus on properties of the coupled light-matter system only. . 
Here $Q_\mathrm{tot}:=\sum_i^3 Z_i$ is the total charge and $M:=\sum_i^3 m_i$ the total mass of the three particle system. We note that the canonical variable $\hat{q}_{\alpha}$ and its conjugate momentum $\hat{p}_{\alpha}$ correspond to the displacement field and we have thus mixed the original light and matter degrees of freedom of the Coulomb gauge~\cite{rokaj2018light, Schaefer2020relevance}. Thus physical observables of the photon field, e.g., the transverse electric field fluctuations, can depend on the displacement, COM and relative coordinates (see, e.g., Eq.~\eqref{eq:Efluc}). The resulting stationary eigenvalue problem can be solved numerically exactly using the wave function ansatz
 \begin{eqnarray}
 \ket{\psi^\prime_{\bold{k},n}} &=& e^{\iu \bold{k}\cdot\boldsymbol{R}_c}\ket{\Phi^\prime_{k_z,n}},\label{eq:ansatz}
 \end{eqnarray} 
where we have  chosen the cavity mode polarized along $z$, wave vectors $\bold{k}$ and quantum numbers $n$. The solution can be achieved by a smart choice of a spherical-cylindrical coordinate system, where angular integrals are treated analytically and radial integrals numerically by using Gauss-Laguerre quadrature.\cite{hesse1999lagrange,hesse2001lagrange,sidler2020chemistry}
From the choice of our gauge an interesting property of the Hamiltonian becomes immediately evident for charged molecules with $Q_\mathrm{tot}\neq0$, e.g. HD$^+$. For those molecules, the COM motion directly couples to the displacement field of the cavity,\cite{sidler2020chemistry} which will add additional numerical complexity to our subsequent numerical treatment in thermal equilibrium. 

% Eq Hamiltonian
\subsection{Thermal Equilibrium in Polaritonic Systems \label{ch:thermeq}}
% quantum statistical description a priori open problem for mixed bosonic-fermionic systems.
The rigorous quantum statistical treatment of a hybrid light-matter system poses interesting theoretical questions, since it contains bosonic and fermionic degrees of freedom that are strongly mixed (we note here that the nuclear degrees of freedom can be both, fermionic or bosonic, depending on the effective spin of the nuclei). In the general case the canonical equilibrium density operator $\hat{\rho}$ is a stationary solution of the quantum Liouville equation,
\begin{eqnarray}
[\hat{H},\hat{\rho}]\overset{!}{=}0,\label{eq:quantumliou}
\end{eqnarray}
subject to the constraints of constant particle number, volume and temperature. The canonical density operator assumes the following general form at temperature $T$,
\begin{eqnarray}
\hat{\rho}
&=&\sum_{n} \frac{e^{-\beta E_{n}}}{\mathcal{Q}} \ket{\psi_{n}}\bra{\psi_{n}}, \label{eq:candens}%\\
%
%&\overset{N=1}{=}&\sum_{\bold{k},n} \frac{e^{-\beta E_{\bold{k},n}}}{\mathcal{Q}}\ket{\psi^\prime_{\bold{k},n}}\bra{\psi^\prime_{\bold{k},n}},
\end{eqnarray}
where $E_n$ and $\ket{\Psi_n}$ are the corresponding $N_p$-particle eigenenergies and eigenstates, $\mathcal{Q}:=\sum_{n} e^{-\beta E_{n}} $ the canonical partition function and $\beta=1/(k_B T)$. In the traditional uncoupled case, i.e., for $\lambda_\alpha=0$, where the $N_p$ fundamental particles in Eq.~\eqref{eq:PFHeigfull} could for example form $N$ spatially distinct molecules (assuming dilute limit), we can simplify the problem by means of statistical physics. Hence, we can treat these $N$ molecular entities either as effective bosons or fermions, i.e. we can occupy the new quasi-particle states according to a fermionic or bosonic statistics. In more detail, we can thermally populate the corresponding $N$-particle states $\ket{\psi_n} \approx \ket{\psi_{n_1}^{(1)}} \wedge ... \wedge \ket{\psi_{n_N}^{(1)}}$ for effective fermions, or $\ket{\psi_n} \approx \ket{\psi_{n_1}^{(1)}} \odot ... \odot \ket{\psi_{n_N}^{(1)}}$  for effective bosons with $E_n \approx E_{n_1}^{(1)} + ... + E_{n_N}^{(1)}$. Here we have introduced the single-molecule eigenenergies $E_{n}^{(1)}$ and eigenstates $\ket{\psi^{(1)}_n}$. For the uncoupled case $\lambda_\alpha =0$, the bare photon modes $\alpha$ just obey the usual Bose-Einstein distribution.  Consequently, the thermal density matrix operator of Eq.~\eqref{eq:candens} would just be a tensor product of the thermal density matrix of the (non-interacting) molecules and the uncoupled photon modes. 

In the strong coupling case $\lambda_{\alpha}>0$ things become complicated. In this case, this simple tensor product ansatz might, however, be no longer sufficient, since the matter and photon degrees of freedom can strongly mix and we \textit{apriori} loose a clear entity to treat statistically (e.g. spatially separated molecules). Indeed, the assumption (sometimes employed in polaritonic chemistry) that light and matter can form a quantum-coherent "super-molecule" inside a cavity~\cite{galego2015cavity,galego2017many} would suggest to treat the complete ensemble of molecules plus cavity as a single quantum entity. If this were the case also for higher temperatures, we would have a macroscopic quantum state at ambient conditions with potential quantum entanglement between the cavity and the ensemble of molecules, which seems rather implausible. Moreover, in this case the fundamental quantum statistics of the individual $N_p$ particles as used in Eq.~\eqref{eq:candens} might become dominant and we need to consider the individual particles completely delocalized over macroscopic distances at ambient conditions. While the rigorous quantum treatment of such an ensemble of molecules is numerically not feasible, we can investigate thermal quantum properties (e.g. light-matter entanglement) for the simplest case $N=1$, i.e., we just have a single HD$^+$ molecule strongly coupled to the cavity. In this case we will have access to the exact thermal density matrix of Eq.~\eqref{eq:candens} since we can calculate the lowest-lying (ro-vibrational) eigenstates of Eq.~\eqref{eq:ansatz}. The numerical details of our approach are described in the Supporting Information of Ref.~\citenum{sidler2020chemistry} as well as in App.~\ref{app:numsol} below with focus on the thermal quantum ensembles.

A few remarks: It is important to contrast the above notion of chemical systems being quantum-coherently coupled with other type of effective quantum models for excitations, e.g., polariton-excitons. In these situations, it is not the wave function of the ensemble of molecules that is being considered, but the excitation's quasi-particle instead, i.e., merely the quantized excitation are being transferred between fixed molecular structures.  We further note that in the case of variable particle numbers, the statistical grand-canonical ensemble should represent different realizations of $N_p$ particles coupled to a cavity as opposed to an indefinite (Fock-space) number of particles coupled to a single cavity. However, the grand-canonical treatment will not be discussed further in this work.

\section{Exact Quantum Properties for Vibrational Strong Coupling at Finite Temperature}

Having the exact thermal equilibrium density operator available for a real molecule under vibrational strong coupling conditions enables to approach the questions raised in the introduction. In the following we will see how the strong light-matter coupling condition induces finite (!) temperatures for the matter and light subsystems, despite keeping the total system temperature at 0 Kelvin, and how the subsystem temperatures approach the canonical temperature when we couple the cavity-molecule system to an external heat bath. Next we will investigate the effect of the hybridization between light and matter on thermal and quantum fluctuations. Finally we discuss quantum entanglement between light and matter at cryogenic temperatures yet show how increasing the temperature destroys quantum entanglement, contrary to predictions from the ubiquitous Jaynes-Cummings model. This resulting classical nature of light-matter interaction makes the appearance of a  quantum coherent ``super-molecule'' implausible at ambient conditions.

\subsection{Temperature Under Strong Coupling Conditions}

% Observables / Definition of subsystem temperature (reduced system) representation vs. full non-equilibrium. JC like vs non-JC results (Show subsystem temperature interpretation, Deviation in observables, show frequency dependency of fluctuations (cv))

%Having numerically exact densities available for a polariton in canonical equilibrium, it is tempting to investigate and quantify cavity induced non-equilibrium effects onto its subsystem constituent. If significant, they will be of fundamental importance for the future developement of polaritonic rate theories under vibrational strong coupling.\cite{sidler2021perspective}
Having numerically exact canonical ensemble densities available at temperature $T$, it is interesting to investigate how the strong coupling condition affects the separate molecular and photon subsystem temperature. As discussed above, the presence of the strongly coupled cavity mode breaks the common weak coupling assumption for the matter subsystem, which will lead to a non-canonical thermal subsystem density matrix operator.
In the following, we will quantify the cavity-induced temperature effects on the matter and light subsystem levels for vibrational strong coupling.  
%
%For our purpose, we  of a reduced canonical density operator  
%\begin{eqnarray}
%\hat{\rho}_{W}:=\sum_l \frac{e^{-\frac{E_{l}}{k_B \tau_W}}}{\mathcal{Q}_W}\ket{l}\bra{l}
%\end{eqnarray} 
%for subsystem $W$ at temperature $\tau_W$, where the full polaritonic system can be partitioned into $W\otimes V$ obeying Eq. (\ref{eq:candens}). This implicitely assumes that the subsystem's dynamics shall be described by a unknown (!) Hamiltonian operator $\hat{H}_W$ with eigenvalues $E_{l}$ and eigenfunctions $\ket{l}$ that is weakly coupled to a bath, i.e. is in canonical equilibrium at temperature $\tau_W$. 
%\revDS{Details missing how to determine?}
%

For this purpose, we introduce a natural definition of subsystem temperatures $\tau$ in terms of the reduced density matrix (RDM) formalism, which will provide access to subsystem-equilibrium properties, given that the full light-matter system is in canonical equilibrium at temperature $T$. Naturally, the definition of a subsystem-temperature $\tau_W$ for a strongly coupled subsystem $W$ involves some ambiguities as we will see, except for the weak coupling limit $\lambda_\alpha\rightarrow 0$, where one should recover canonical properties for the subsystem $W$, i.e. $\tau_W\rightarrow T$. We also note the obvious connection to quantum embedding schemes such as subsystem density-functional theory~\cite{https://doi.org/10.1002/wcms.1175} or density-matrix embedding theory~\cite{doi:10.1021/ct301044e,PhysRevB.101.075131} %Moreover, our RDM approach provides a measure for the non-equilibrium effects, which emerge for certain observables due to the strong light-matter coupling.
Let the RDM operator of $\hat{\rho}$ given in Eq.~\eqref{eq:candens} be defined by
\begin{eqnarray}
\hat{\rho}_{W,\overline{V}}:=Tr_V[\hat{\rho}],
\end{eqnarray}
for a bi-partite partitioning of the full polaritonic system $W\otimes V$.
The bar indicates the traced out vector space $V$. It is straightforward to show that $\hat{\rho}_{W,\overline{V}}$ remains a self-adjoint operator with  $Tr_W[\hat{\rho}_{W,\overline{V}}]=1$, due to the normalisation of the ensemble density matrix operator $\hat{\rho}_{W,V}$ by the canonical partition function $\mathcal{Q}$.
Because $\hat{\rho}_{W,\overline{V}}$ is self-adjoint on the Hilbert space $W$ we have a unique diagonal representation
\begin{align}
 \hat{\rho}_{W,\overline{V}} = \sum_{l} w_l \ket{l}\bra{l}.   
\end{align}
This always allows to define, for an arbitrary (!) temperature $\tau_{\rm arb}$, a self-adjoint operator for which $\hat{\rho}_{W,\overline{V}}$ represents a canonical ensemble. We can do so by choosing $E_{l}^{\rm arb}$ such that
\begin{align}
    e^{-\tfrac{E_l^{\rm arb}}{k_b \tau_{\rm arb}}} = w_l,
\end{align}
which leads to $\hat{H}_{W}^{\rm arb} = \sum_{l} E_{l}^{\rm arb} \ket{l}\bra{l}$. So to find a physically reasonable definition of a subsystem temperature we need to fix the subsystem Hamiltonian $\hat{H}_W$. In the case of coupled light-matter systems this can be done naturally by taking $\lambda =0$ in Eq.~\eqref{eq:HDplusHamiltonian} and considering the decoupled light and matter Hamiltonian. In this case we can further subdivide the matter Hamiltonian in COM and relative matter system, i.e, we use the notation $W \in \{{\rm pt, COM, m}\}$ for the different subsystems. Using the corresponding subsystem Hamiltonians we can then determine
\begin{eqnarray}
E_{l}^{W}=\bra{l}\hat{H}_W\ket{l}
\end{eqnarray} 
and then get numerically the corresponding subsystem temperature $\tau_{W}$ by the fitting
\begin{align}
\inf_{\tau_W} \left(\sum_{l} \left|w_l - \exp(-\tfrac{E_{l}^{W}}{k_b \tau_{W}}) \right|\right).    
\end{align}
We expect this choice to be reasonable for moderately interacting subsystems, which remain close to an equilibrium at $\tau_W\approx T$. Obviously, our subsystem temperature definition will not work anymore for very strongly interacting subsystems (e.g. light and matter under ultra-strong coupling conditions, or the partitioning of a molecular bond into two pieces). In that case, there would be no reason to expect an exponential fitting for $\tau_W$ to be reasonable and the non-canonical contributions would be dominating. We note that by construction, for non-interacting subsystem $W$ and $V$ (e.g. photon mode and matter at $\lambda_{\alpha}=0$), the RDM operators equal the subsystem canonical density matrix operators $\hat{\rho}_{W}$, i.e., $\Tr_V\hat{\rho}=\hat{\rho}_{W}$ with $T=\tau$. This is a common assumption for weakly interacting subsystems of interest, which considerably reduces the complexity of the problem, but cannot be imposed under strong vibrational coupling conditions, as we will show subsequently.%, that we put to the test.
Let us first consider the simple COM subsystem. For the temperature of the COM motion one immediately finds
\begin{eqnarray}
\tau_\mathrm{COM}=T,
\end{eqnarray}
because the eigenfunctions  of our fully coupled HD$^+$ Hamiltonian given in Eq.~\eqref{eq:ansatz} ensures that the full Hamiltonian and ensemble density matrix operator are blockdiagonal with respect to the quantum numbers $k$. Therefore, the partial trace operation acting on the relative matter and photonic degrees of freedom reduces each block to one dimension. Consequently, both reduced matrices are diagonal, which trivially obey $[\hat{H}_{\mathrm{COM}},\hat{\rho}_{\mathrm{COM}}]=0$. This implies that the COM dynamics obeys strict canonical equilibrium within the long-wavelength limit of the Pauli-Fierz theory. This is a nice consistency between the classical idea of the temperature of a gas, which assumes a certain distribution of velocities of particles, and the quantum-mechanical treatment.

\begin{figure*}
     \centering
     \begin{subfigure}{0.45\textwidth}
         %\caption{}%$y=x$}
        \centering
         \includegraphics[width=82mm]{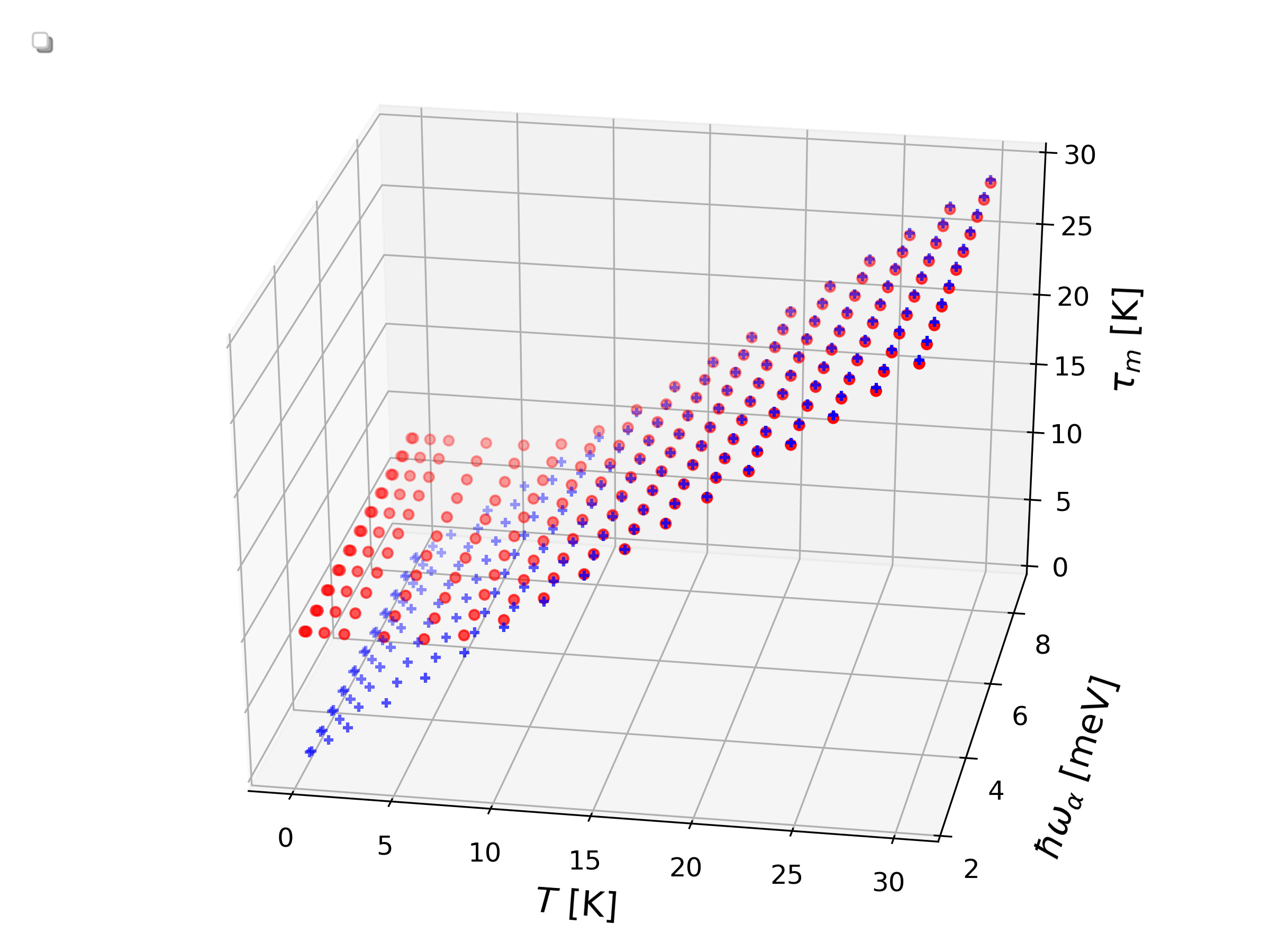}
        %\label{fig:tau_m}
     \end{subfigure}%
     \begin{subfigure}{0.45\textwidth}
         %\caption{}%$y=x$}
        \centering
         \includegraphics[width=82mm]{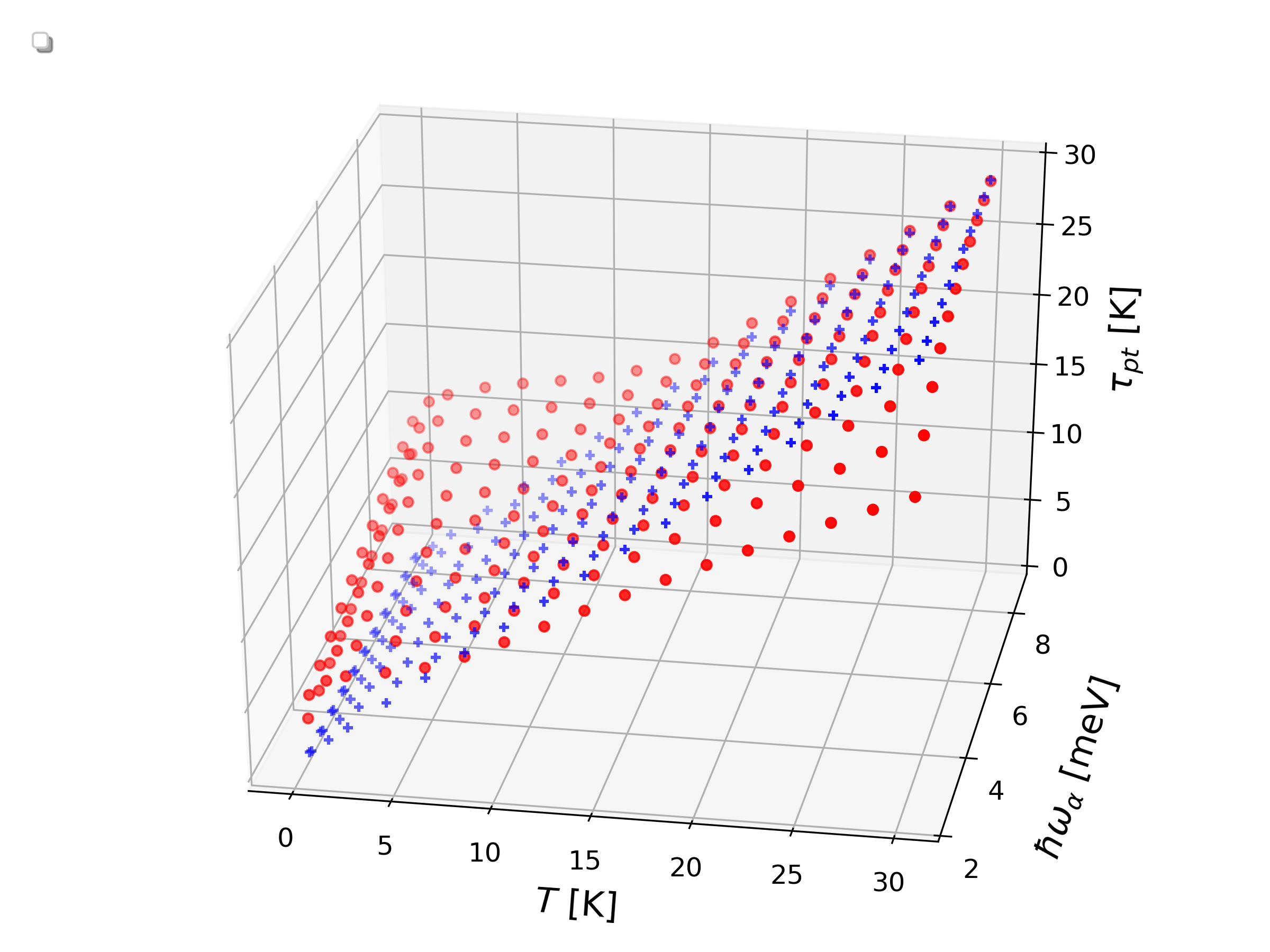}
        %\label{fig:tau_pt}
     \end{subfigure}%
        \caption{Left: Heating and cooling effects for the matter subsystem temperatures $\tau_{\rm m}$ emerging from cavity-induced non-equilibrium conditions under vibrational strong coupling for $\lambda_{\alpha}=0.005$ at $\hbar\omega_\alpha$ for different total system temperatures $T$. Red dots correspond to the exact solution for HD$^+$ coupled to a single cavity mode, whereas blue dotes visualize the equilibrium value for the weak coupling assumption, i.e. $\tau=T$.
        Right: The same analysis for cavity-induced modifications of the photon mode temperature $\tau_{\rm pt}$ under vibrational strong coupling.}
        \label{fig:tau}
\end{figure*}

However things change fundamentally for the relative matter temperature $\tau_{\rm m}(T,\lambda_{\alpha},\omega_{\alpha})$ and photon temperatures $\tau_{\rm pt}(T,\lambda_{\alpha},\omega_{\alpha})$ as displayed in Fig.~\ref{fig:tau} for ro-vibrational strong coupling with $\lambda=0.005$ [a.u.] for frequencies close to the first ro-vibrational excitation of HD$^+$ at $\omega=5.4$ meV. Notice that the aforementioned blockdiagonal nature of the full ensemble density matrix significantly simplifies the numerics of those calculations, because the partial trace operation acting on the COM subsystem then effectively reduces to the trace operator summing over $k_z$. Our temperature definitions already indicates that the strong light-matter coupling induces different heating as well as cooling effects on the subsystems, which depend on the fixed temperature $T$ and coupling $\lambda_{\alpha}$ and partially on the  cavity mode frequency $\omega_{\alpha}$. In more detail, we observe two different regimes for the matter temperature $\tau_{\rm m}$. It converges to a finite minimal matter temperature for $T\lessapprox T_0=10$ K. This lower bound for the matter temperature $\tau_{\rm m}(T\rightarrow 0,\lambda_{\alpha}>0, \omega_{\alpha})>0$ and the corresponding transition temperature $T_0$ strongly depends on the chosen light-matter coupling $\lambda_{\alpha}$ and virtually no dependency on the chosen resonance frequency $\omega_{\alpha}$ was observed. As we will see also in the following sections, at approximately $T_0$ not only the matter subsystem temperature starts to deviate strongly from the externally defined (canonical) temperature, but also other important properties of the coupled light matter system change their character. The transition temperature $T_0$ is therefore a characteristic quantity of the coupled HD+ system. For $T_0 \lessapprox T $, we find that $\tau_{\rm m}\lessapprox T$, i.e. it almost corresponds to the temperature of the total system with a slight cooling involved.
For the temperature of the strongly coupled photon mode $\tau_{\rm pt}$ we find a different behaviour. Still at low $T\lessapprox T_0$ there is a clear heating observed, i.e. $\tau_{\rm pt}>T$. However, this turns into a significant cooling $\tau_{\rm pt}<T$ for larger $T$. In contrast to the relative matter subsystem, the magnitude of the heating and cooling regimes strongly depends on the chosen cavity frequency $\omega_{\alpha}$. The qualitative difference of $\tau_{\rm pt}$ and $\tau_{\rm m}$ is not surprising, since HD$^+$ is a charged molecule. Therefore, the thermal COM motion $k_z\neq 0$ along the polarization  $ \boldsymbol{\epsilon}_\alpha \parallel \boldsymbol{k}_z$  will significantly affect the photon-field, i.e. the thermal center of charge motion is formally equivalent to the pumping of the cavity with external currents. \textit{A priori} this ``temperature pumping'' effect will directly increase the photon number and thus affect $\tau_{\rm pt}$, but much less (i.e. only indirectly) the relative molecular system. In contrast, we would expect qualitative similar behaviour for $\tau_{\rm pt}$ and $\tau_{\rm m}$ for a neutral molecule under vibrational strong coupling, i.e. (virtually) no dependency on $\omega_{\alpha}$.

We want to highlight that at ultra-low temperatures $T\approx 0$ the vibrational strong coupling seems to induce a non-equilibrium condition for the subsystems, which can be regarded as a significant heating, i.e. the absence of $0$ K for matter and light. This finding may be particularly relevant for the future interpretation of experimental data in the low cryogenic regime, e.g. for modifications of the critical temperature of cavity assisted superconductivity\cite{thomas2019exploring} and other polaritonic phenomena (condensates) at low $T$.  % or when trying to reach long decoherence times for the entangled states in cavity assisted quantum computing \revDS{is there any reference for this?}. \textcolor{red}{Need to explain in which way and why. Forgot the arguments that we came up with.} 

\subsection{Cavity Modified Thermal Fluctuations}

In a next step, we investigate how the strong ro-vibrational coupling affects the (vacuum) field mode and matter fluctuations in thermal equilibrium. Reaching a detailed understanding of cavity-modified fluctuations is not only of theoretical interest, but it is of fundamental importance for the emerging fields of polaritonic chemistry and materials science, where modified fluctuations would call for an adaptation of usual molecular dynamics simulations and rate theories~\cite{sidler2021perspective}. For example, changing the dynamics (fluctuations) of matter in a (cavity)-frequency selective manner under thermal equilibrium conditions opens new pathways to steer and control chemical reactions,.\cite{thomas2019tilting} %modify material properties\cite{?} and develop novel devices \revDS{e.g. ??}.\cite{?}

For this reason, we subsequently investigate the exact field and matter dipole fluctuations accessible for our HD$^+$ molecule under ro-vibrational strong coupling.
As previously stated, the strongly coupled HD$^+$ molecule is diagonalised in the COM- relative length gauge which follows from the transformation given in Eq.~\eqref{eq:pzw}. Consequently, to obtain physically meaningful results, we need to also transform the usual Coulomb-gauged observables to our gauge choice. That is, when evaluating the respective physical observables $\hat{O}$ defined in (the velocity form of the) Coulomb gauge (see Eq.~\eqref{eq:PFHeigfull}), we consider $\hat{O}^\prime:= S\circ \hat{U}\hat{O}\hat{U}^\dagger$ instead.  

%Accordingly, the pt-number operator in the Coulomb gauge $\hat{n}_{pt,\alpha}=\hat{a}^\dagger_\alpha \hat{a}_\alpha$ of the coupled mode $\alpha$ transforms to,
%\begin{eqnarray}
%\hat{n}_{pt,\alpha}^\prime&=& 
%\hat{a}_{\alpha}^\dagger \hat{a}_{\alpha}+\lambda_\alpha\sum_{i=1}^3 Z_i \hat{z}_{ci} \hat{q}_\alpha+\frac{1}{2 \omega_\alpha}\Big(\lambda_\alpha\sum_{i=1}^3 Z_i \hat{z}_{ci}\Big)^2.
%\label{eq:numpt}
%\end{eqnarray}
%
 We find the transformed vector potential $\boldsymbol{\hat{A}}^\prime$, the displacement field $\boldsymbol{\hat{D}}^\prime$ and the transverse electric-field $\boldsymbol{\hat{E}}^\prime$ operators, polarized along the polarization axis of the cavity $z$ as 
\begin{eqnarray}
 \hat{A}_z^\prime&=& \frac{\lambda_\alpha \hat{p}}{\omega_\alpha} ,\\
 \hat{D}^\prime_z&=&  \lambda_\alpha\omega_\alpha\hat{q}_\alpha,\\
\hat{E}^\prime_z&=&  \hat{D}^\prime-\lambda_\alpha^2\hat{d}_z \label{eq:eprime} .
%
%\langle \hat{E}_z^2\rangle-\langle \hat{E}_z\rangle^2&=&\sum_{k_z,n} \bra{\Phi^\prime_{k_z,n}}(\lambda_\alpha\omega_\alpha\hat{q}_\alpha)^2+\lambda_\alpha^4\sum_{i=1}^3 Z_i^2 \hat{z}_{ci}^2 \ket{\Phi^\prime_{k_z,n}} \frac{e^{-\beta \hat{E_{k_z,n}}}}{\mathcal{Z}_{\mathrm{min}}}\nonumber\\
%&&-\bigg(\sum_{k_z,n} \bra{\Phi^\prime_{k_z,n}}\lambda_\alpha\omega_\alpha\hat{q}_\alpha \ket{\Phi^\prime_{k_z,n}} \frac{e^{-\beta \hat{E_{k_z,n}}}}{\mathcal{Z}_{\mathrm{min}}}\bigg)^2-
%\sum_{i=1}^3\bigg(\sum_{k_z,n} \bra{\Phi^\prime_{k_z,n}}\lambda_\alpha^2 Z_i \hat{z}_{ci} \ket{\Phi^\prime_{k_z,n}} \frac{e^{-\beta \hat{E_{k_z,n}}}}{\mathcal{Z}_{\mathrm{min}}}\bigg)^2\\
%&=&\Delta(\lambda_\alpha\omega_\alpha\hat{q}_\alpha)+\sum_{i=1}^3\Delta(\lambda_\alpha^2 Z_i \hat{z}_{ci})
\end{eqnarray}
The respective squared operators for the field and dipole fluctuations then become
\begin{eqnarray}
\hat{A}_z^{\prime 2}&=&\Big(\frac{\lambda_\alpha \hat{p}_\alpha}{\omega_\alpha} \Big)^2,\label{eq:afluc}\\
\hat{D}_z^{\prime 2}&=&(\lambda_\alpha\omega_\alpha\hat{q}_\alpha)^2,\\
\hat{E}_z^{\prime 2}&=&\hat{D}^{\prime 2}_z-2\hat{D}^\prime_z\lambda_\alpha^2\hat{d}_z+\lambda_\alpha^4\hat{d}_z^2\label{eq:Efluc}\\
\hat{d}_z^2&=&\Big(\sum_{i=1}^3 Z_i \hat{z}_{ci}\Big)^2\label{eq:dfluc}
%
%\langle \hat{E}_z^2\rangle-\langle \hat{E}_z\rangle^2&=&\sum_{k_z,n} \bra{\Phi^\prime_{k_z,n}}(\lambda_\alpha\omega_\alpha\hat{q}_\alpha)^2+\lambda_\alpha^4\sum_{i=1}^3 Z_i^2 \hat{z}_{ci}^2 \ket{\Phi^\prime_{k_z,n}} \frac{e^{-\beta \hat{E_{k_z,n}}}}{\mathcal{Z}_{\mathrm{min}}}\nonumber\\
%&&-\bigg(\sum_{k_z,n} \bra{\Phi^\prime_{k_z,n}}\lambda_\alpha\omega_\alpha\hat{q}_\alpha \ket{\Phi^\prime_{k_z,n}} \frac{e^{-\beta \hat{E_{k_z,n}}}}{\mathcal{Z}_{\mathrm{min}}}\bigg)^2-
%\sum_{i=1}^3\bigg(\sum_{k_z,n} \bra{\Phi^\prime_{k_z,n}}\lambda_\alpha^2 Z_i \hat{z}_{ci} \ket{\Phi^\prime_{k_z,n}} \frac{e^{-\beta \hat{E_{k_z,n}}}}{\mathcal{Z}_{\mathrm{min}}}\bigg)^2\\
%&=&\Delta(\lambda_\alpha\omega_\alpha\hat{q}_\alpha)+\sum_{i=1}^3\Delta(\lambda_\alpha^2 Z_i \hat{z}_{ci})
\end{eqnarray}
Notice that the physical transverse electric field operator corresponds to the displacement field operator in the standard velocity form of the Hamiltonian operator given in Eq.~\eqref{eq:PFHeigfull}, i.e. $\boldsymbol{\hat{E}}=\boldsymbol{\hat{D}}^\prime$. However, our specific gauge choice introduced the dependency on the relative dipole operator as given in Eq.~\eqref{eq:eprime}. 

The conservation of the parity symmetry $P$ for the Hamiltonian operator $\hat{H}$ as well as for $\hat{H}^\prime$ in our COM-relative gauge has interesting consequences for the fully quantized system. From the Hamiltonian invariance under $(\hat{\boldsymbol{r}},\hat{q}_\alpha)\overset{P}{\mapsto}(-\hat{\boldsymbol{r}},-\hat{q}_\alpha)$ a zero transversal field and zero dipole condition follows: 
\begin{eqnarray}
\langle \boldsymbol{\hat{E}}^\prime\rangle_{\bold{k},n}&=&\langle \boldsymbol{\hat{A}}^\prime\rangle_{\bold{k},n}=\langle
\boldsymbol{\hat{D}}^\prime\rangle_{\bold{k},n}=\langle \boldsymbol{\hat{d}}\rangle_{\bold{k},n}=\langle\boldsymbol{\hat{r}}_{ci}\rangle_{\bold{k},n}=0.\label{eq:zerocomponent}
\end{eqnarray}
This implies that we need to break the parity symmetry of the Hamiltonian in order to have a finite molecular dipole, e.g. by fixing the nuclei on a Born-Oppenheimer surface. In this way we choose a specific realization of the otherwise symmetric possibilities of free space~\cite{anderson1972more}. In practice the choice of which possibility is realized is then governed by the local environment. This has interesting consequences for a potential "super-molecule" of a quantum-coherent ensemble of molecules, as we will discuss later. %\revDS{is this statement correct? be more precise}
%Interestingly, for the fully quantized system, no direction is distinguished. In more detail (proof?), one can show that the dipole and displacement field expectation values are componentwise zero \revDS{is this really true?}:
Consequently, photon field and matter dipole fluctuations of the form $\Delta O=\langle \hat{O}^2\rangle-\langle\hat{O}\rangle^2$ can entirely be described by Eqs.~\eqref{eq:afluc}-\eqref{eq:dfluc}. Notice the Eq.~\eqref{eq:zerocomponent} allows to disentangle thermal E-field fluctuations $\langle\hat{E}_z^{\prime 2}\rangle_T$ in terms of dipole $\langle\hat{d}_z^2\rangle_T$ and displacement-field fluctuations $\langle\hat{D}_z^{\prime 2}\rangle_T$ as well as their respective quantum correlations following from $\langle\hat{D}_z^\prime\hat{d}_z\rangle_T\neq 0$,  which become exactly zero in the classical limit due to the previous parity argument, i.e. from Eq.~\eqref{eq:zerocomponent}. The magnitude of this gauge-dependent (!) quantum correlations is \textit{a priori} of no physical interest. However, it becomes a relevant quantity for the future development of  approximations in theoretical models or for simulation methods under cavity-modified thermal equilibrium conditions (e.g. in terms of open quantum systems\cite{sidler2021perspective} or within semi-classical cavity Born-Oppenheimer molecular dynamics~\cite{flick2017cavity,sidler2021perspective}).

\begin{figure}
     \centering
     \begin{subfigure}{0.45\textwidth}
         %\caption{}%$y=x$}
        \centering
         \includegraphics[width=65mm]{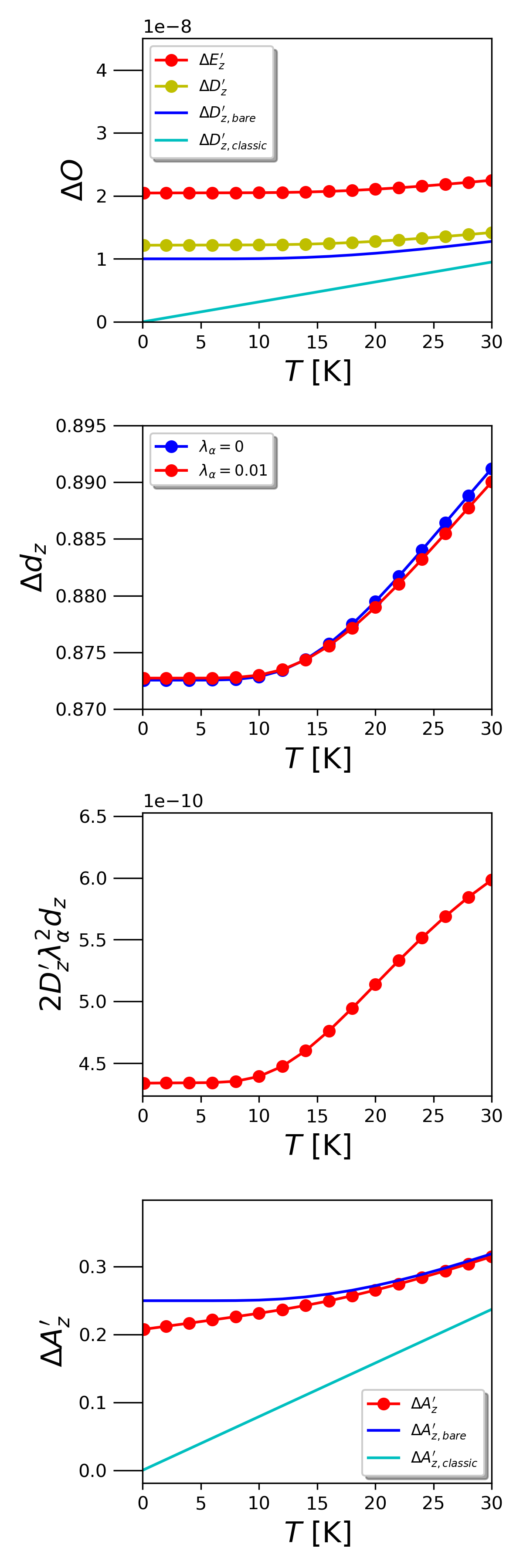}
     \end{subfigure}%
        \caption{From top to bottom: Thermal quantum fluctuations for electric $\Delta E_z^\prime$ and displacement field  $\Delta D_z^\prime$, dipole $\Delta d_z$, dipole-displacement correlations $2D_z^\prime\lambda_\alpha ^2 d_z$ and vector potential $\Delta A_z^\prime$ operators. Comparing to the uncoupled fluctuations ($\lambda=0$) reveals two different fluctuation regimes bellow and above $T^0\approx 15$ K for a coupling strength of $\lambda=0.01$ and the cavity tuned on the first ro-vibrational excitation of HD$^+$. }
         \label{fig:fluct}
\end{figure}

When tuning the cavity on resonance with the first ro-vibrational excitation of HD$^+$, we find the temperature dependency of the fluctuations as shown in Fig.~\ref{fig:fluct} for $\lambda_{\alpha}=0.01$. When evaluating the ensemble averages of the operators given in Eqs.~\eqref{eq:afluc}-\eqref{eq:dfluc}, we observe a significant increase (shift) of the transverse electric field fluctuations $\langle\hat{E}_z^{\prime 2}\rangle_{T}$ compared with thermal vacuum fluctuation $\langle\hat{D}_z^{\prime 2}\rangle_{T,\ \rm bare}$  of a bare cavity mode, due to the strong coupling with matter. The displayed analytical electric/displacement field fluctuations of a bare cavity mode can be calculated analytically as
\begin{eqnarray}
\langle\hat{D}_z^{\prime 2}\rangle_{T,\ \rm bare} &=& \langle\hat{E}_z^{\prime 2}\rangle_{T,\ \rm bare}=\lambda_{\alpha}^2 \omega_{\alpha} \Bigg[ \frac{1}{2}+\frac{e^{-\frac{\omega_{\alpha}}{k_B T}}}{1-e^{-\frac{\omega_{\alpha}}{k_B T}}}\Bigg],\label{eq:barefluc}
\end{eqnarray}
which converges to 
\begin{eqnarray}
\langle\hat{D}_z^{\prime 2}\rangle_{T,\ \rm classic}&\overset{k_B T\gg\hbar\omega_{\alpha}}{\longrightarrow}&\lambda_\alpha^2 k_B T,\label{eq:clasfluc}
\end{eqnarray}
in the classical limit for $k_B T\gg\hbar\omega_{\alpha}$.
%We can relate the dressed thermal field fluctuations, which are available only numerically to the analytic solution of the uncoupled photon mode, i.e. for a bare quantum harmonic oscillator of frequency $\omega_{\alpha}$.
While the dressed electric field fluctuation overall are shifted to higher values, the temperature dependency remains more or less preserved with respect to the thermal quantum fluctuations of a bare cavity mode. For the thermal matter fluctuations of the coupled dipole operator, i.e. for $\langle\hat{d}_z^2\rangle_T$, we find a slight suppression at temperatures $T< T^0\approx 15$ K with $\lambda_\alpha=0.01$, followed by an increase of the fluctuations at higher temperatures, which indicates the transition to a different regime of physics at  a temperature $T^0$, which is in agreement with the previous observations for the subsystem temperatures. 
Similarly, the gauge-dependent light-matter quantum correlations of the form $2\hat{D}_z^\prime\lambda_\alpha^2\hat{d}_z$  change from a slight increase to a small suppression. However, overall they remain negligibly small, i.e. two orders of magnitude smaller than the physically relevant transverse electric field fluctuations. Consequently, quantum correlations between the dressed displacement field and the matter dipole could safely be neglected, which opens room for efficient approximations to investigate more involved systems. In contrast to the increase of the transverse electric fluctuations, our simulation depicts that the thermal fluctuations of the vector potential $\langle\hat{A}_z^{\prime 2}\rangle_{T}$ are suppressed most significantly at low temperatures $T\lessapprox T^0$, compared with a bare cavity mode,
\begin{eqnarray}
\langle\hat{A}_z^{\prime 2}\rangle_{T,\ \rm bare} &=& \frac{\lambda_{\alpha}^4}{ \omega_{\alpha}} \Bigg[ \frac{1}{2}+\frac{e^{-\frac{\omega_{\alpha}}{k_B T}}}{1-e^{-\frac{\omega_{\alpha}}{k_B T}}}\Bigg],
\end{eqnarray}
with its classical counterpart for $k_B T\gg\hbar\omega_{\alpha}$
\begin{eqnarray}
\langle\hat{A}_z^{\prime 2}\rangle_{T,\ \rm classic}&\overset{k_B T\gg\hbar\omega_{\alpha}}{\longrightarrow}&\frac{\lambda_{\alpha}^4}{ \omega_{\alpha}^2} k_B T.
\end{eqnarray}
While beyond $T^0$ the quantum harmonic oscillator solution is quickly approached.

Overall, the theoretically predicted suppression of matter fluctuations $\langle\hat{d}_z^2\rangle_T$ at temperatures far beyond $T^0$ confirm from first principles that the equilibrium dynamics of matter can indeed substantially be modified by ro-vibrational strong coupling to the quantized cavity modes, as proposed in Ref.~\citenum{sidler2021perspective} (before reaching the classical limit $k_B T\gg\hbar\omega_{\alpha}$). This observation has potential impact on the future development of polaritonic reaction rate theories and non-equilibrium simulation methods, which are crucial for the design of novel cavity-mediated reaction processes and for cavity-mediated modifications of the equilibrium ground state in quantum materials. %A rapidly growing field of research which has achieved numerous experimental breakthroughs\cite{?} of industrial relevance so far, but with only limited theoretical understanding~\cite{sidler2021perspective,...}.  
Aside from the significantly modified matter dynamics at high temperatures, the discovered transition to a different fluctuation regime for cryogenic temperatures $T<T^0$ raises the question of the underlying physical mechanism, which we will discuss in the following. 

%%%%%%%%%%%%%%%%%%%%%%%%%%%%%%%%%%%%%%%%%%%%%%%%%%%%%%%%%%%%%%%%%%%%%%%%%%%%%%%%%%%%%%%%%%%%%
\subsection{Cavity Induced Light-Matter Entanglement at Finite Temperature}

Apart from identifying cavity-mediated heating/cooling and correlating thermal fluctuations between light and matter, our exact solution of HD$^+$ in a cavity also allow to assess the ``quantumness'', i.e. the quantum entanglement of the light and matter, at finite temperatures. Entanglement between light and matter would make strongly-coupled molecule-cavity systems for ro-vibrational frequencies interesting for potential applications in quantum-information processing. This would be specifically true if this entanglement would be thermally stable for sizeable temperatures. We further note that to the best of our knowledge, this is the first study that computes light-matter entanglement for a real molecular system in a cavity, i.e. one does not rely on a model Hamiltonian that treats the matter degrees of freedom in a very simplified manner.

To investigate this question we will  determine the temperature-dependent light-matter entanglement under ro-vibrational strong coupling. For this purpose, we rely on the logarithmic negativity 
\begin{eqnarray}
\eta_W(\hat{\rho}):=\log_2\Big(2\sum_{\nu_i<0}|\nu_i|+1\Big),
\end{eqnarray}
which is a computationally efficient (i.e. not $NP$-hard) bipartite entanglement measure applicable to mixed states of distinguishable particles.\cite{peres1996separability,horodecki1996teleportation,eisert1999comparison,vidal2002computable,plenio2005logarithmic}
The negative eigenvalues $\nu_i$ are calculated from the partial transpose of the ensemble density operator $\hat{\rho}^{\Gamma_{W}}$ with respect to the chosen subsystem $W$ in a bipartite partitioning. 
Fortunately, the necessary distinguishability criterion is certainly fulfilled for our dressed HD$^+$ molecule, since its three fermionic constituents are different (electron, proton, deuteron) and couple to one bosonic cavity mode only. 
The logarithmic negativity entanglement measure serves as an upper bound for the distillable entanglement.\cite{plenio2005logarithmic}  However, a zero logarithmic negativity does not imply that the bipartite subsystems are not entangled, since a bound entangled state cannot be detected.\cite{horodecki1998mixed} This has particularly interesting implications for our charged COM motion, which directly couples to the photon field. Indeed the COM motion in a cavity provides a nice example of a bound entangled state with respect the rest of the system. In more detail, we find  
\begin{eqnarray}
\eta_{\rm COM}(\hat{\rho}(T))=0,
\end{eqnarray}
because  of $\hat{\rho}^{\Gamma_{\rm COM}}=\hat{\rho}$, which uses that $\hat{\rho}$ is blockdiagonal with respect to $\boldsymbol{k}$. However, at the same time the COM subsystem is not separable provided that $\lambda_{\alpha}\neq0,\ k_z\neq0,\ \mathcal{Q}_\mathrm{tot}\neq0$. Under these circumstances the charged COM motion along $z$ couples to the photon field, i.e. both factors in the exact eigenfunction  given in Eq.~\eqref{eq:ansatz} depend on $k_z$. Consequently, the COM partition forms a bound entangled pair with the rest of our system.
%(see https://en.wikipedia.org/wiki/Negativity_(quantum_mechanics)#Logarithmic_negativity)
%https://en.wikipedia.org/wiki/Bound_entanglement

\begin{figure}
     \centering
     \begin{subfigure}{0.45\textwidth}
         %\caption{}%$y=x$}
        \centering
         \includegraphics[width=84.5mm]{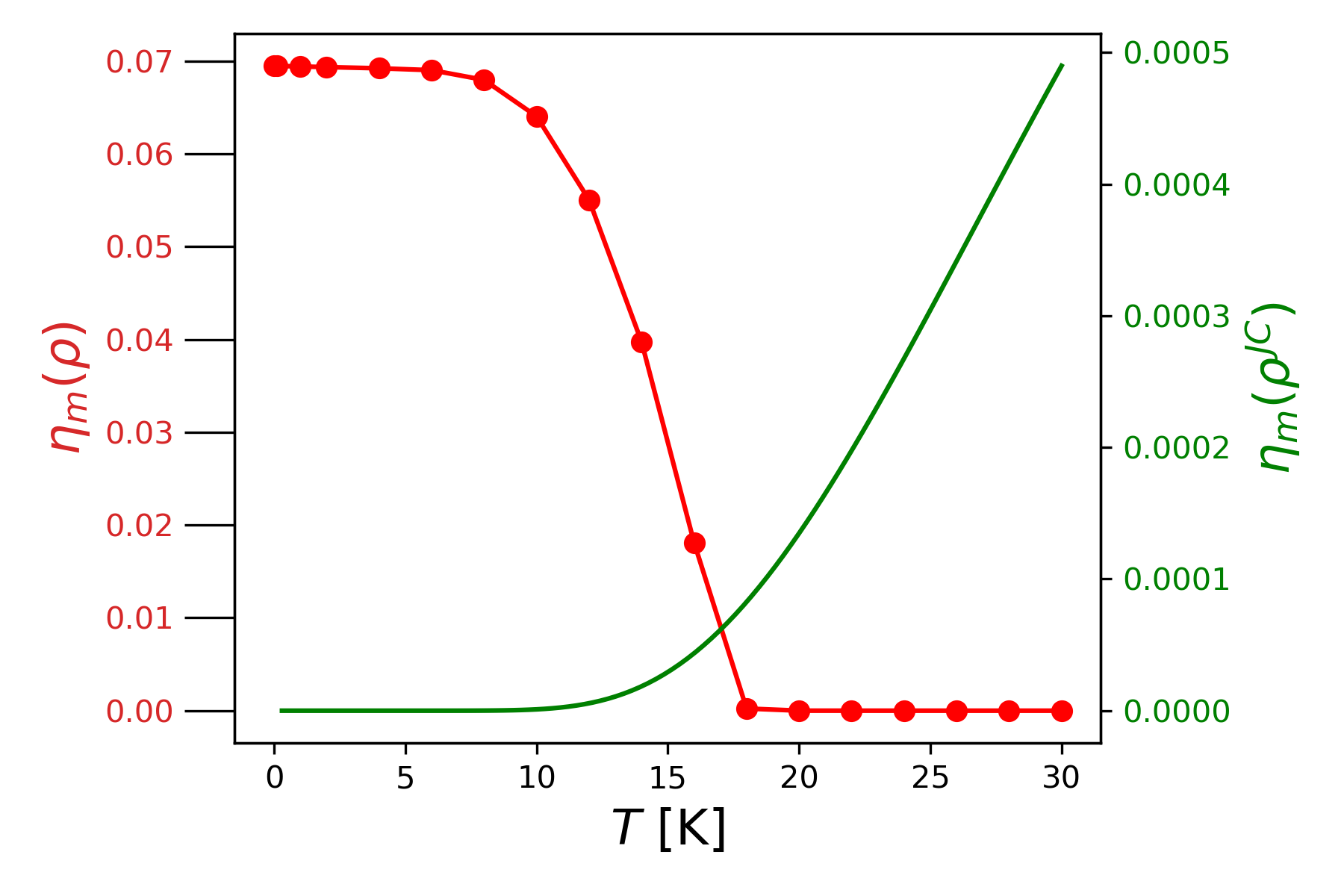}
     \end{subfigure}%
        \caption{Logarithmic negativity measure for the entanglement of light and matter for a HD$^+$ molecule in a cavity under thermal equilibrium conditions. The exact equilibrium solution of the Pauli-Fierz Hamiltonian (red) shows constant entanglement up to $T\approx T^0$, which then is quickly destroyed thermally. In contrast, the phenomenological Jaynes-Cummings model (green) suggest the opposite behaviour, i.e. thermal entanglement creation by mixing the bare matter ground state with excited polaritons. Notice that we can only interpret each measure individually, but we cannot relate the magnitudes for the two systems (acting on different Hilbert spaces), since the logarithmic negativity is not asymptotically continuous.}
        \label{fig:logneg}
\end{figure}

Now let us have a look at the entanglement between light and matter for our dressed HD$^+$ molecule. The detailed numerical procedure to determine the logarithmic negativity 
\begin{eqnarray}
\eta_{\rm m}(\hat{\rho}(T))=\eta_{\rm pt}(\hat{\rho}(T))
\end{eqnarray}
for our system is given in App. \ref{app:numneg}. Because we have already discussed that the coupled COM degrees of freedom cannot contribute to the logarithmic negativity, any non-vanishing value of $\eta_{\rm m}$ can be attributed to entanglement between the relative matter subsystem and the photon field. In Fig.~\ref{fig:logneg} the numerically exact $\eta_{m}$ is displayed (in red) with respect to the temperature $T$, where we have set the coupling to $\lambda_{\alpha}=0.005$ and tuned the cavity on resonance with the first ro-vibrational excitation.
We find significant, almost constant, light-matter entanglement $\eta_m$ between $0$ and $T^0\approx 10$ K, which then quickly drops for higher temperatures and remains zero for temperatures beyond $18$ K.
Consequently, the different physical regime for $T>T^0$ seems to be a consequence of the thermal extinction of the entanglement between light and matter. This indicates that a classical  description for the coupling of light and matter might cover all relevant aspects for temperatures beyond $T^0$. In contrast, the observation of a non-zero logarithmic negativity measure indicates that our hybridized thermal state (dominated by the coupled light-matter ground state) would in principle be suitable for quantum computing in the cryogenic regime $T<T^0$. 
However, not surprisingly, as soon as the temperature becomes large enough such that a sizeable contribution from the excited states is mixed with the  ground state, the light-matter entanglement is lost. By increasing the coupling parameter $\lambda_\alpha$ one can in principle reach entanglement at slightly higher temperatures. However, overall it will be limited by the thermal population of the lowest ro-vibrational excitation, i.e. our simulations confirm that cavity-induced light-matter entanglement for vibrational strong coupling can only be achieved under thermal equilibrium condition at ultra-low temperatures. This effect is a direct consequence of the hybridisation of light and matter in the ground state. Neglecting this delicate aspects of cavity-induced ground state modifications from the Pauli-Fierz Hamiltonian, as commonly done in phenomenological models,\cite{jaynes1963comparison} leads to qualitatively and quantitatively contradicting results.

For example, when applying the logarithmic negativity measure to the ubiquitous Jaynes-Cummings model used to describe polaritonic systems  one finds (see App.~\ref{app:neg_jc}). 
\begin{align}
&\eta_{\rm m}(\hat{\rho}^{JC}(T)) =\nonumber\\
&\log_2\Bigg[1+
\sqrt{e^{-\frac{2E_g}{k_B T}}+\Big(e^{-\frac{E_l}{k_B T}}-e^{-\frac{E_u}{k_B T}}\Big)^2}-e^{-\frac{E_g}{k_B T}}\Bigg],\label{eq:negjc}
\end{align}
assuming a cavity tuned on resonance with the bare matter excitation. %For simplicity, only the bare matter ground-state energy $E_g$ and the energy eigenvalues of the first polaritonic branch composed of the lower  $E_l$ and upper $E_u$ polariton energy were considered respectively. 
The Jaynes-Cummings model implies a bare matter ground-state and thus automatically leads to
\begin{eqnarray}
\eta_{\rm m}(\hat{\rho}^{JC}(T\rightarrow 0))=0,
\end{eqnarray}
%and cannot be compensated by considering counter-rotating terms or higher lying polaritonic states.\revDS{Ruggi: is this correct? I.e. can counter rotating terms modify the groundstate? of the couple system, or is it purly excited state theory?}
Indeed, neglecting the hybrid light-matter nature of the ground state introduces a thermal light-matter entanglement as shown in Fig.~\ref{fig:logneg} in green, i.e. the thermal mixing of the eigenstates creates entanglement under equilibrium conditions instead of (correctly) destroying it with increasing temperatures.
Notice, in our work we investigate light-matter entanglement under thermal equilibrium conditions. Clearly, when preparing the system initially in an excited polaritonic state (e.g. lower or upper polariton) things will change and the Jaynes-Cummings model may become a reasonable approximation again. %In this case, the light-matter entanglement will mostly be limited by finite lifetime, i.e. cavity losses and decoherence effects and it  strongly depend on the quality of the initial preparation.\revDS{more precise / specific?}
In contrast, the ground-state-dominated light-matter entanglement at thermal equilibrium corresponds to a stationary solution of the system (see Eq.~\eqref{eq:quantumliou}), which at sufficiently low temperatures is a long-lived and robust with respect to perturbations/decoherence effects. We note that even including the full continuum of modes of the electromagnetic field, i.e. radiative dissipation, will keep the ground state of the molecule infinitely lived and thus a true bound state in the continuum which is completely decoherence free~\cite{spohn2004}.

\section{Discussion, Conclusion and Outlook}

\begin{figure*}
     \centering
     \begin{subfigure}{1\textwidth}
        \centering
         \includegraphics[width=160mm]{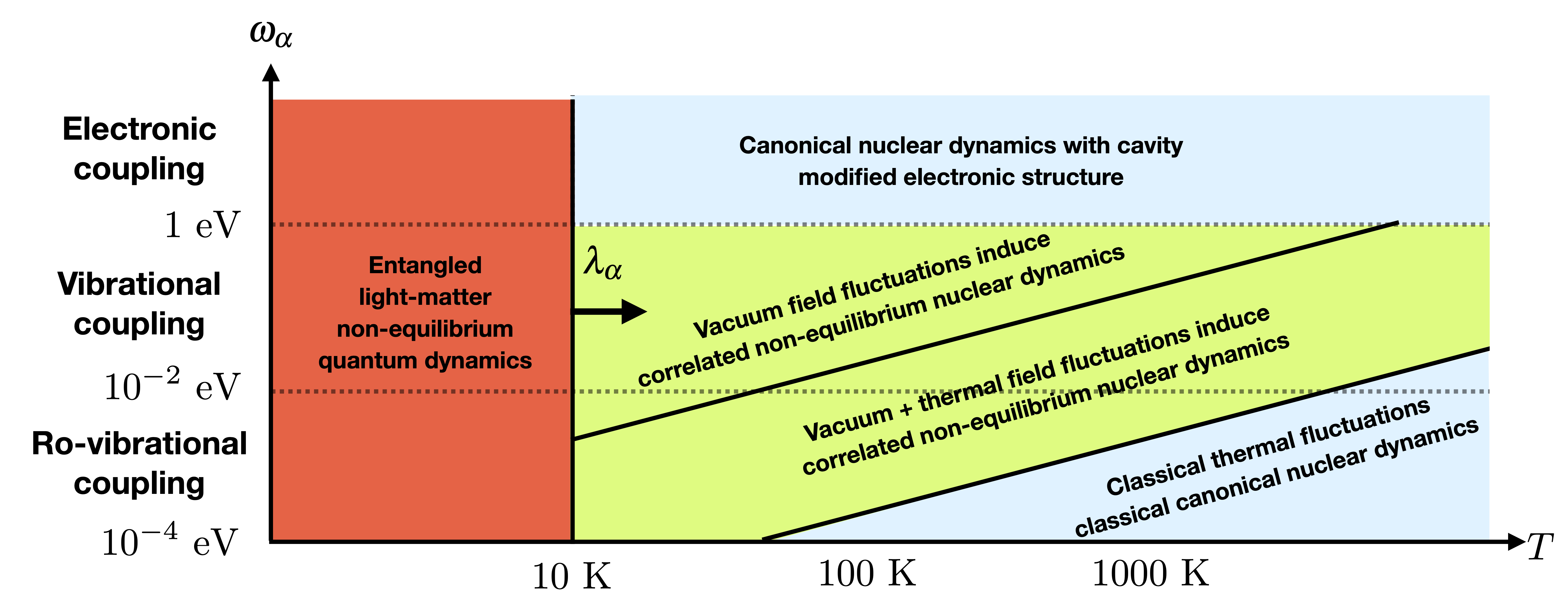}
     \end{subfigure}%
      \caption{Pictorial sketch of different thermal (non)-equilibrium regimes emergent under molecular strong coupling conditions in a cavity. The interpretation was inferred from the exact simulation results for one HD$^+$ molecule coupled to a single cavity mode.  }%$y=x$}
       \label{fig:overview}
\end{figure*}

Having investigated the role of temperature, quantum fluctuations and entanglement for the HD$^+$ under ro-vibrational strong coupling to a single cavity photon mode, we can collect our findings to develop a concise picture of polaritonic systems in thermal equilibrium. Despite being limited to exact results of the HD$^+$ molecule, strongly coupled in relatively narrow frequency regime at low temperatures, we can extrapolate fundamental dynamical characteristics from this prototypical molecular polaritonic system to a broader frequency and temperature range. Thanks to the exact first principle approach, the number of initial assumptions and approximations could be reduced to a minimum, which provides unprecedented insights as summarized in Fig.~\ref{fig:overview}.  
Based on our simulation results, we suggest to distinguish three different regimes, where the (nuclear) dynamics is governed by different type of physics:
\begin{enumerate}[(i)]
    \item First, at low cryogenic temperatures $T<T^0$ of the combined system, we find \textbf{light-matter entanglement} $\eta(\lambda_\alpha)>0$, which arises from cavity-induced modifications of the ground state. Consequently, an accurate theoretical description requires \textit{a priori} the full quantum treatment of light and matter. This automatically implies cavity induced \textbf{non-equilibrium quantum dynamics} for the respective subsystems in absence of external driving. 
    Furthermore, having distillable quantum entangled states available in the ground state of molecular polaritonic systems, may also be of interest for the design of robust entangled states suitable for quantum computing.\cite{yu2016long,wasielewski2020exploiting,kavokin2022polariton}
    Note also that the discovered heating of the subsystem temperatures due to strong light-matter interaction, effectively prevents the sub-systems from reaching 0 K, despite approaching the hybridised groundstate of the total system at 0 K.  This effect should potentially be considered, for the interpretation of experimental results in the ultra low cryogenic regime, e.g.  for cavity mediated superconductors. 
    \item By increasing the system's temperature $T$, thermal mixing of eigenstates quickly destroys the quantum entanglement between light and matter at $T>T^0$ even in the strong coupling regime. Consequently, we enter the regime of \textbf{correlated light-matter dynamics} (see also Fig. S1 in the Supporting Information for IR to visible strong coupling regimes). %Similar to then entangled case, we find the transversal electric field fluctuations  significantly increased compared with the bare cavity mode, whereas the dipole fluctuations of matter are considerably suppressed after loosing the light-matter quantum entanglement. Notice also that the significant displacement field fluctuations $\Delta D$ are still a consequence of the quantum nature of the fields, where two different mechanisms add up.
    However,  the field fluctuations are still governed by quantum laws influencing the matter via strong coupling, even in absence of light-matter entanglement.
    We can distinguish two sub-cases: 
    \begin{enumerate}
        \item At low thermal energies, i.e. $T^0< T\leq \hbar\omega_\alpha/k_B$, the disentangled field fluctuations are mainly driven by the vacuum fluctuations of the (dressed) ground state of the hybrid light-matter system. Overall, the coupling to matter enhances the fluctuations compared with a bare cavity mode (i.e. in our setup the coupled transversal electric (vacuum) field fluctuations are doubled).
        \item At moderately higher temperatures $\hbar\omega_\alpha/k_B\lessapprox T $, thermal mixing of few excited states start to contribute  to the field fluctuations. Therefore, a quantum thermal description is still required, before reaching the classical thermal limit for $k_B T\gg\hbar\omega_\alpha$.
    \end{enumerate}
    %From the computationally accessible observables, we cannot decide if the accompanying suppression of the matter fluctuations  are a consequence of the potential energy modification under single molecular strong coupling, which results in the squeezing of the nuclear distances (see Fig. S? in the Supporting Information) or  if it is also driven by the coupling to the thermal field fluctuations. %, which results in an over overall suppression of the energy fluctuations (see Fig. S? in the Supporting Information\revDS{specific heat}).
   Interestingly, the correlation of quantum fluctuations (vacuum and thermal) with matter can be sustained well beyond room-temperature for typical molecular coupling frequencies in the vibrational and ro-vibrational regimes, as we can infer from the analytic solution of the bare mode fluctuations (i.e., by comparing Eqs. (\ref{eq:barefluc}) \& (\ref{eq:clasfluc})).  This has particularly interesting consequences for the theoretical description of chemical processes in a polarionic setting.  
   First of all, the observed loss of light-matter entanglement at low cryogenic temperatures opens the door for a semi-classical description/interpretation of the cavity-induced (non)-equilibrium physics. For example in terms of a cavity Born-Oppenheimer picture,\cite{flick2017cavity} which treats the nuclei and photons classically, but accounts for the non-classical nature of the field fluctuations.\cite{sidler2021perspective} %which suggests that the matter fluctuation are indeed altered by a combination of potential energy modifications and by the quantum nature of the field fluctuations, i.e. $\Delta d_z(\Delta D,\lambda_\alpha,T)$.
   % In more detail, a classical description for the photons and the nuclei should in principle reasonably approximate the most relevant physical processes in the regime of correlated light-matter dynamics (e.g. cavity Born-Oppenheimer (CBO) approximation\cite{flick2017cavity}), provided that the non-classical nature of the field fluctuation is still accounted for.\cite{sidler2021perspective} %As shown in Fig. \ref{fig:fluct}, the thermal and vacuum displacement-field fluctuations can for example relatively well be approximated by a bare quantum harmonic oscillator mode in thermal equilibrium, which couples classically to the matter, i.e. the involved quantum correlations could safely be neglected. 
    %
    Choosing this semi-classical picture to interpret the physical mechanisms has two interesting consequences. First, it can trivially be generalised to polaritonic ensembles with different matter constituents, e.g. $N>1$, since the classical nuclei and photons become distinguishable, which avoids the initially stated issues of mixed quantum-statistics or the partitioning into non-interacting subsystems. It also agrees with the usual approach to molecular ensembles where the large amount of molecules allows a semi-classical statistical description. For instance, for a large ensemble of gas phase molecules no permanent dipole appears as the orientations of the dipoles fluctuate randomly~\cite{anderson1972more}.
    Second, the coexistence of classical and quantum fluctuations in a Langevin-setting\cite{sidler2021perspective} at room temperature is expected to induce stochastic resonances in the coupled system characterized by \textbf{``classical'' non-equilibrium nuclear dynamics}. Stochastic resonances have been proposed as a mechanism to explain the experimentally observed resonance phenomena in cavity-mediated chemical reactions,\cite{sidler2021perspective} which emerge in absence of external periodic driving under ambient conditions,\cite{Thomas2016,thomas2019exploring} but have not yet been fully rationalised theoretically. % may emerge,  \ref{sidler2021perspective,to two temperature simulations, stoch res} correlating vibrations over large distances for sufficiently complex matter systems (e.g. between different molecular bonds).\cite{schafer2021shining} which gives rise to \textbf{``classical'' non-equilibrium nuclear dynamics}, due to the strong coupling of classical nuclei  to the quantum nature of the dressed thermal and vacuum mode fluctuations.\cite{sidler2021perspective} %assuming a coupled system of classical nuclei exposed to classical collisions (fluctuations) with a collective dipole coupling to the classical cavity mode, governed by quantum fluctuations, eventually introduces non-canonical equilibrium dynamics.\cite{sidler2021perspective}
    %As a consequence, cavity   Clearly, our investigated HD+ molecule provides too few nuclear degrees of freedom to correlate different bond vibrations as for example done in the semi-classical setting of Ref.\citenum{schafer2021shining}.  
    %However, our simulations confirm that the regime of correlated light-matter dynamics is indeed of high practical relevance for ground state chemical applications, since light-matter entanglement is already lost at low cryogenic temperatures, but the induced non-equilibrium / stochastic resonance mechanisms are expected to reach far beyond room-temperature, depending on the chosen cavity frequencies $\omega_\alpha$.
    %
    \item In the high temperature limit $k_B T\gg \hbar \omega_\alpha$, the thermal fluctuations of the cavity  eventually approach the classical limit, which suggests that we reach  \textbf{classical canonical nuclear dynamics}\cite{sidler2021perspective} and aforementioned stochastic resonance phenomena can no longer emerge.\cite{sidler2021perspective} Therefore, modifications of ground state chemical processes will most likely be dominated by canonical free energy modifications induced in a cavity. For example, single molecular strong coupling is believed to be induced/enhanced in a collectively coupled molecular environment,  as a consequence of local perturbations (e.g. nuclear displacement).\cite{sidler2020polaritonic} Furthermore, cavity modified non-adiabatic effects from excited electronic states may also start to play a significant role. 
    
    Similarly to the high temperature limit, we anticipate canonical nuclear dynamics also in the high frequency limit (electronic strong coupling), which cannot be accessed with our numerical setup, however. 
    In more detail, when coupling the cavity to electronic excitation energies %different physical mechanism may start to play a role in thermal equilibrium, which we cannot infer from our HD+ molecular simulations. However, for \textbf{electronic strong coupling} conditions 
    we expect that thermal field fluctuations will be negligible under ambient conditions and thus only the vacuum mode fluctuations will alter the electronic properties locally\cite{sidler2020polaritonic} as well as collectively. Therefore, the dynamics of the nuclei will most likely be well described by pure classical canonical dynamics in thermal equilibrium, except at ultra-low temperatures, where (as always) quantum statistics starts to play a significant role. This picture is further supported by recent simulation results, which confirm that nuclear dynamics under electronic strong coupling is well described classically in a Born-Oppenheimer picture with separated electronic and photonic degrees of freedom.\cite{flick2017cavity}   % Consequently, nuclear properties for electronic strong coupling will be dominated by modifications of the potential energy surface, as well as non-adiabatic / dark state effects  may start play a significant role (e.g. in photochemistry). 
    Whether or not electronic strong coupling introduces entanglement between light and matter beyond the cryogenic temperature regime can not be deduced from our simulations, but it seems rather unlikely. 
\end{enumerate}

The above-developed theoretical picture of light matter interaction unifies our exact findings with previous theoretical considerations on polaritonic chemistry in Ref.~\citenum{sidler2021perspective} and approximate simulation results in Ref.~\citenum{schafer2021shining,sidler2020polaritonic} as well as experimental evidence in Refs.~\citenum{Thomas2016,thomas2019tilting}. This confirms that necessary prerequisites are fulfilled to tackle polaritonic chemistry under vibrational strong coupling by means of stochastic resonance phenomena, which highlights a new research direction in chemistry, i.e. the development of stochastic resonance driven (non-equilibrium) rate theories for chemical reactions. 
Moreover, the discovered absence of zero temperature for strongly coupled matter (and light) subsystems, is considered to be highly relevant for ultra cold polaritonic systems, e.g. for quantum computing or super conducting experiments.  Especially, since it seems to be strongly related to the emergence and loss of light-matter entanglement.
However, clearly our theoretical picture of cavity-modified (non)-equilibrium dynamics of realistic molecular systems is still sketchy at the moment and requires substantial future refinement. In the following, we briefly address two relevant research directions that we plan to pursue next based on the findings of the present work:

\begin{enumerate}
    \item \textit{Mixed Quantum Statistics:} As described in Section ~\ref{ch:thermeq}, we still lack a quantum-statistical equilibrium description of (indistinguishable) matter, which is strongly coupled to photonic modes, for two reasons: First, the strong coupling of the quantized modes to the collective matter dipole \textit{a priori} hinders the partitioning of the ensemble into weakly interacting entities. Second, if a certain partitioning is assumed, the theoretical treatment of mixed bosonic/fermionic particle statistics has only been marginally explored so far.\cite{buchholz2019reduced,buchholz2020light,buchholz2021many} Having a thorough quantum-statistical description available will be of pivotal relevance for the better understanding of molecular polaritonic phases in the ultra-low temperature regime, where light-matter entanglement might play a significant role. A detailed understanding of how entanglement is build up in this regime and how mixed quantum statistics might help to protect such entanglement also for higher temperatures is an interesting question to provide robust entangled states. %, and to assess our semi-classical interpretation for temperatures above $T^0$. 
    A promising starting point in this direction could be an open quantum systems setting. %i.e. performing exact time-propagation including dissipation effects which then might converge to some (non-canonical) stationary solution for the subsystems or remain in non-equilibrium. 
    However, standard open quantum systems methods (e.g. Gorini-Kossakowski-Sudarshan-Lindblad formalism\cite{lindblad1976generators,gorini1976completely,head2020quantum}) are typically restricted to the dilute gas limit, assuming non-interacting bosonic or distinguishable molecular entities as well as weak coupling to an external bath, i.e. they usually impose Markovian dynamics.\cite{head2020quantum} Those assumptions are often not met in practice for realistic molecular systems (e.g. liquids) under strong vibrational coupling conditions, where non-Markovian processes become important, as we have demonstrated in this work. Recently, there have been extensions introduced for non-Markovian dynamics\cite{head2019ensemble,head2019satisfying} and fermionic systems\cite{nguyen2015electronic,head2015communication}, which may help to gain a detailed theoretical understanding of the dynamics of entangled or correlated polaritonic systems with mixed quantum statistics.

    \item \textit{The Impact of Collective Effects:} The presence of multiple (identical) molecules $N>1$ can significantly enhance the coupling strength of light and matter, which is most prominently identified by a $\sqrt{N}$-scaling behaviour of the Rabi-splitting in the Dicke model.\cite{dicke1954coherence} While the general relevance of collective scaling-effects on various observables in polaritonic systems is undisputed, i.e. they appear in observables that probe the entire system (e.g. optical absorption or non-linear spectroscopy), little is known on how the collective coupling translates into the individual single molecular light-matter coupling.. While phenomenological models suggest quantum properties at room temperature by construction,\cite{galego2017many} semi-classical polarizability (e.g. radiation reaction approach) may in principle be sufficient to capture collective effects on chemical reactions at ambient conditions.\cite{schafer2022shortcut,schafer2022polaritonic} Such an approach would agree with the usual semi-classical interpretation of molecular ensembles. For a quantum-coherent "super-molecule" the question of symmetry of the total ensemble would become important. For instance, quantum-mechanically small molecules have no permanent dipole since one finds the various non-symmetric (permanent dipole) solutions super-imposed and only the environment favors one over the other. On the other hand, for large molecules or ensembles the symmetries are only obeyed statistically~\cite{anderson1972more}. That is, to switch quantum-mechanically between different symmetry states of large molecules becomes more and more unlikely with an increase of the system size. A "super-molecule" could lift such quantum-mechanical switching to macroscopic scales. Our exact results cannot address such collective aspect, since we are limited to $N=1$. However, the exact loss of light-matter entanglement at low cryogenic temperatures makes the quantum nature of collective strong coupling effects highly questionable at ambient conditions. Significant future research effort is needed for a better understanding and description of quantum and classical collective effects in polaritonic molecular ensembles. The possibility to use collectivity at ultra-low temperatures to enhance light-matter entanglement is, however, intriguing. Here the interesting connection to ultra-cold chemistry seems worthwhile to explore further~\cite{doi:10.1126/science.aam6299}.  
    
    %\item \textit{Dark State Ensembles:}
    %Apart from the quantum / classical nature of collective effects, the hybridisation of light and matter in principle creates collective dark states,\cite{?} which are believed to significantly alter thermal ensemble properties. The presence  of highly degenerate dark states significantly modifies the density of states in a thermal ensemble for $N\gg 1$, which should modify thermal equilibrium properties of polaritonic ensembles. However, to what extend this mechanism affects chemical properties at ambient conditions, remains to be explored under vibrational stong coupling if one considers our findings of non-canonical nuclear dynamics, as well as the lack of a proper quantum statistical description for $N>1$, which accounts for the presence of fermionic matter. 
 
\end{enumerate}
Overall, we think that our exact thermal simulation results of a real molecular system under vibrational strong coupling conditions, paves the way for many future theoretical and experimental work in various research disciplines not only in quantum physics, but also in chemistry and materials science in general.  

%\section{Conclusion and Outlook}

\begin{acknowledgments}
This work was made possible through the support of the RouTe Project (13N14839), financed by the Federal Ministry of Education and Research (Bundesministerium für Bildung und Forschung (BMBF)) and supported by the European Research Council (ERC-2015-AdG694097), the Cluster of Excellence “CUI: Advanced Imaging of Matter” of the Deutsche Forschungsgemeinschaft (DFG), EXC 2056, project ID 390715994 and the Grupos Consolidados (IT1453-22). 
The Flatiron Institute is a division of the Simons Foundation.
\end{acknowledgments}

\section*{Author Contributions}
D.S. initiated the project, developed the simulation code and the required theoretical concepts, suitable for numerical calculations. M.R. initiated the subsystem temperature concept. 
All authors developed the physical interpretation and wrote the manuscript.

\section*{Data Availability Statement}
Numerical data available upon request.

\bibliographystyle{unsrt}
\bibliography{manuscript}

\appendix

\section{Numerical Solution of Pauli-Fierz Hamiltonian and Ensemble Averaging\label{app:numsol}}

The exact solution of our quantized 3-body matter system coupled to a cavity mode relies on the choice of a highly optimized, i.e. problem specific, coordinate representation,\cite{hesse1999lagrange,hesse2001lagrange} in combination with Gauss-Laguerre numerical quadrature (see Supporting Information of Ref. \citenum{sidler2020chemistry} for  numerical details and the Supporting Information of this work for specific computational parameter choices).

When evaluating ensemble averages numerically, we assume $V\rightarrow\infty$, which implies a continuum of quantum numbers $k_z$, which allows an efficient numerical evaluation, i.e. we replace $\sum_{k_z}\mapsto\int_\infty^\infty dk_z$. Hence, we can apply the Gauss-Hermite quadrature procedure to approximate the infinite integral along $k_z$ numerically, i.e. by using the following relation,\cite{shizgal1981gaussian}

\begin{eqnarray}
\int_{-\infty}^\infty \frac{1}{\sigma \sqrt{2\pi}}e^{-\frac{(y-\mu)^2}{2\sigma^2}} h(y) dy \approx \frac{1}{\sqrt{\pi}}\sum_{i=1}^s w_i h(\sqrt{2}\sigma x_i+\mu),
\end{eqnarray}
where $x_i$ are the roots of the Hermite polynomial $H_s(x)$ and the weights $w_i=\frac{2^{s-1} s!\sqrt{\pi}}{s^2 [H_{s-1}(x_i)]^2}$.

In our case, we map the ensemble averages of observable $\hat{A}$ as,
\begin{eqnarray}
\langle \hat{A}\rangle_T& =& \int_\infty^\infty dk_z \bigg(\sum_n\frac{e^{-\beta E_{k_z,n}}}{\mathcal{Z}_{\mathrm{min}}} \bra{\Phi^\prime_{k_z,n}}\hat{A}^\prime\ket{\Phi^\prime_{k_z,n}}\bigg) \nonumber\\
&=&\int_\infty^\infty dk_z e^{-\frac{\beta}{2M} \big(k_z-\frac{\lambda_\alpha Q_{\mathrm{tot}}\langle\hat{p}_\alpha\rangle_{0}}{\omega_\alpha}\big)^2}\\
&&\bigg(\sum_n\frac{e^{-\beta E_{k_z,n}^{\mathrm{red}}}}{\mathcal{Z}_{\mathrm{min}}} \bra{\Phi^\prime_{k_z,n}}\hat{A}^\prime\ket{\Phi^\prime_{k_z,n}}\bigg) \nonumber\\
&=&\int_\infty^\infty dk_z e^{-\frac{\beta}{2M} \big(k_z-\frac{\lambda_\alpha Q_{\mathrm{tot}}\langle\hat{p}_\alpha\rangle_{0}}{\omega_\alpha}\big)^2}h(k_z) \nonumber\\
&\approx&\sqrt\frac{2 M}{\beta}\sum_{i=1}^s w_i h\bigg(\sqrt{\frac{2 M}{\beta}}  x_i+\frac{\lambda_\alpha Q_{\mathrm{tot}}\langle\hat{p}_\alpha\rangle_{gs}}{\omega_\alpha}\bigg)
\end{eqnarray}
and accordingly,
\begin{eqnarray}
\mathcal{Z}_{\mathrm{min}}&=&\int_\infty^\infty dk_z e^{-\frac{\beta}{2M} \big(k_z-\frac{\lambda_\alpha Q_{\mathrm{tot}}\langle\hat{p}_\alpha\rangle_{0}}{\omega_\alpha}\big)^2}\bigg(\sum_n e^{-\beta E_{k_z,n}^{\mathrm{red}}}\bigg)\\
&=&\int_\infty^\infty dk_z e^{-\frac{\beta}{2M} \big(k_z-\frac{\lambda_\alpha Q_{\mathrm{tot}}\langle\hat{p}_\alpha\rangle_{0}}{\omega_\alpha}\big)^2}g(k_z)\\
&\approx&\sqrt\frac{2 M}{\beta}\sum_{i=1}^s w_i g\bigg(\sqrt{\frac{2 M}{\beta}}  x_i+\frac{\lambda_\alpha Q_{\mathrm{tot}}\langle\hat{p}_\alpha\rangle_{gs}}{\omega_\alpha}\bigg).
\end{eqnarray}
This assumes that the transformed $\hat{A}^\prime$ does not depend on $k_z,\ k_y,\ \hat{R}_{c}$ and $\hat{P}_c$, which is indeed the case for all our observables. The COM reduced Energy eigenvalue is defined as $E_{k_z,n}^{\mathrm{red}}:=E_{k_z,n}-\big(k_z-\frac{\lambda_\alpha Q_{\mathrm{tot}}\langle\hat{p}_\alpha\rangle}{\omega_\alpha}\big)^2$. Hence, for the determination of $\mu=\frac{\lambda_\alpha Q_{\mathrm{tot}}\langle\hat{p}_\alpha\rangle_{0}}{\omega_\alpha}$ it is assumed that the ground-state contribution $\langle \hat p_\alpha\rangle_{0}$ to the ensemble average dominates. However, from the symmetry argument in Eq. (\ref{eq:zerocomponent}) we immediately notice  $\mu=0$, since $\langle \hat p_\alpha\rangle=0$ for all states in absence of external driving currents, which makes accurate integration possible for few discrete $k_z$-evaluations only.
%In case of no coupling to external driving currents, as we assume throughout this work, we can further simplify our calculations as follows. By employing the cavity Born-Oppenheimer approximation\cite{flick2017cavity} $\hat{p}_\alpha\mapsto p_\alpha$, which allows to apply the Hellmann-Feynmann theorem for $\{k_{xy},k_z, p_\alpha \}$, which then states that for the ground-state energy
%\begin{eqnarray}
%\frac{d\langle H^\prime\rangle_0}{d k_{xy}}&=& \frac{k_{xy}}{M}=0,\\
%\frac{d\langle H^\prime\rangle_0}{d k_z}&=& %\frac{k_z}{M}+\frac{Q_{\mathrm{tot}}\lambda_\alpha}{M\omega_\alpha} p_\alpha=0,\\
%
%\frac{d\langle H^\prime\rangle_0}{d p_\alpha}&=
%& \Big(\frac{Q_{\mathrm{tot}}k_z\lambda_\alpha}{M\omega_\alpha}\Big)+
%\bigg[\frac{1}{M}\Big(\frac{\lambda_\alpha Q_{\mathrm{tot}}}{\omega_\alpha} \Big) ^2 +1\bigg]p_\alpha=0.\label{eq:dhdpalpha}
%\end{eqnarray}
%
%Setting the derivatives to zero, gives access to one unique extremal values of the parametrized potential energy surface for,
%\begin{eqnarray}
%k_z^\mathrm{min}&:=&-\frac{Q_{\mathrm{tot}}\lambda_\alpha}{\omega_\alpha} p_\alpha^\mathrm{min}\overset{Eq. \ (\ref{eq:dhdpalpha})}=0\label{eq:kz_extimate}, %\\
%p_\alpha^\mathrm{min}&=& j_\alpha \sqrt{\frac{2}{\omega_\alpha}},\\
%p_{\alpha^\prime}^\mathrm{min}&=&0
%\label{eq:minkz}
%\end{eqnarray}
%i.e.  the ground-state COM remains zero for a charged molecule within a cavity, as it is the case in free space and thus $\mu =0$ can savely be assumed for the numerical integrals over $k_z$.

%%%%%%%%%%%%%%%%%%%%%%

\section{Numerics for Logarithmic Negativity\label{app:numneg}}

%
%how to define for continuous case? i.e. for $\bold{R}_c$??? not so clear! use that $e^{ikR}e^{-ikR}=1$ vanishes I guess $\sum_i w_i = \int d^3_k e^{\beta E_k}/Z$. Transpose should not change anything because of plane wave for COM? What does this mean for entanglement with COM? 
The canonical density matrix given in Eq. (\ref{eq:candens}) can be written for our system in terms of COM, relative and photon basis explicitly as,
\begin{eqnarray}
\hat{\rho}&=&\sum_{\bold{k},n}\frac{e^{-\beta E_{\bold{k},n}}}{\mathcal{Z}} \sum_{i,s}\sum_{j,t}c^*_{i,s}(k_z,n)c_{j,t}(k_z,n)\nonumber\\
&&\ket{e^{-\iu \bold{k R}_c}}\ket{\bold{r}_{ci}}_i\ket{q_\alpha}_s\bra{e^{\iu \bold{k R}_c}}\bra{\bold{r}_{ci}}_j\bra{q_\alpha}_t,\label{eq:rhobasis}
\end{eqnarray}
which accounts for the blockdiagonal nature in terms of COM coordinates.
For our subsytem choices $W\in\big\{ \bold{r}_{ci}, \hat{q}_\alpha\big\}$ the partial transpose follows from Eq. (\ref{eq:rhobasis}) by either $c^*_{i,s}c_{j,t}\mapsto c^*_{j,s}c_{i,t}$ or $c^*_{i,s}c_{j,t}\mapsto c^*_{i,t}c_{j,s}$, leading e.g. to the partial transpose of the photon subsystem,

\begin{eqnarray}
\hat{\rho}^{\prime\Gamma_{q_\alpha}}_{\alpha}&=&\sum_{\bold{k},n}\frac{e^{-\beta E_{\bold{k},n}}}{\mathcal{Z}} \sum_{i,s}\sum_{j,t}c^*_{i,t}(k_z,n)c_{j,s}(k_z,n)\nonumber\\
&&\ket{e^{-\iu \bold{k R}_c}}\ket{\bold{r}_{ci}}_i\ket{q_\alpha}_s\bra{e^{\iu \bold{k R}_c}}\bra{\bold{r}_{ci}}_j\bra{q_\alpha}_t,
\end{eqnarray}
which can be diagonalised yielding the corresponding eigenvalues $w^{\Gamma_{q_{\alpha}}}_{\bold{k}^\prime,l}(T)=\frac{e^{-\beta \frac{k_x^{\prime 2}+k_y^{\prime 2}+k_z^{\prime 2}}{2M}}}{\mathcal{Z}}w^{\Gamma_{q_{\alpha}}}_{k_z^\prime,l}$ and eigenfunctions $\ket{e^{-\iu \bold{k^\prime R}_c}}\sum_{i,s} d_{i,s}^{\Gamma_{q_{\alpha}}}(k_z^\prime,l)\ket{\bold{r}_{ci}}_i\ket{q_\alpha}_s$. For a fixed $k_z$-value, this diagonalization can efficiently be computed numerically for our HD$^+$ molecule and the choice of our subsystems.  
The logarithmic negativity entanglement measure then follows immediatly from:
\begin{eqnarray}
\eta_W(\hat{\rho})&=&\log_2\Bigg(\sum_{\bold{k},l}|w^{\Gamma_{W}}_{\bold{k}^\prime,l}|-w^{\Gamma_{W}}_{\bold{k}^\prime,l}+1\Bigg)\\
&=&\log_2\Bigg(\int dk^3 \sum_l|w^{\Gamma_{W}}_{\bold{k}^\prime,l}|-w^{\Gamma_{W}}_{\bold{k}^\prime,l}+1\Bigg),
\end{eqnarray}
where in the last step the summation over $\bold{k}$ were  approximated by an integral, as it was previously the case for the ensemble averages. 

\section{Jaynes-Cummings Light-Matter Entanglement in Canonical Equilibrium\label{app:neg_jc}}

The three energetically lowest  eigenfunctions of the JC-model are 
\begin{eqnarray}
\ket{\Psi_g}&=&\ket{g}\otimes\ket{0}\\
\ket{\Psi_l}&=&\frac{1}{\sqrt{2}}(\ket{e}\otimes\ket{0}-\ket{g}\otimes\ket{1})\\
\ket{\Psi_u}&=&\frac{1}{\sqrt{2}}(\ket{e}\otimes\ket{0}+\ket{g}\otimes\ket{1})
\end{eqnarray}
which are composed from the bare matter $\{\ket{g},\ket{e}\}$ and photon  $\{\ket{0},\ket{1}\}$ eigenstates, assuming the rotating wave approximation. Furthermore, we have assumed for simplicity that the cavity is tuned exactly on resonance with the first bare matter excitation energy $\omega_{\alpha}=E_e-E_g$.
Expressing the corresponding canonical equilibrium density  in the light-matter basis $\{\ket{g}\otimes\ket{0},\ \ket{g}\otimes\ket{1},\ \ket{e}\otimes\ket{0},\ \ket{e}\otimes\ket{1}\}$  leads to the following matrix representation,
\begin{eqnarray}
\rho^{JC}&=& 
\begin{bmatrix}
e^{-\frac{E_g}{k_B T}} & 0 & 0 & 0 \\
0 & \frac{e^{-\frac{E_l}{k_B T}}}{2} + \frac{e^{-\frac{E_u}{k_B T}}}{2}& -\frac{e^{-\frac{E_l}{k_B T}}}{2}+\frac{e^{-\frac{E_u}{k_B T}}}{2} & 0 \\
0 & -\frac{e^{-\frac{E_l}{k_B T}}}{2}+\frac{e^{-\frac{E_u}{k_B T}}}{2} & \frac{e^{-\frac{E_l}{k_B T}}}{2}+\frac{e^{-\frac{E_u}{k_B T}}}{2} & 0 \\
0 & 0 & 0 & 0 
\end{bmatrix}  
\end{eqnarray}
with partial transpose
\begin{widetext}
\begin{eqnarray}
\rho^{\Gamma_{m},JC}=&&\nonumber\\
&\begin{bmatrix}
e^{-\frac{E_g}{k_B T}} & 0 & 0 & -\frac{e^{-\frac{E_l}{k_B T}}}{2}+\frac{e^{-\frac{E_u}{k_B T}}}{2} \\
0 & \frac{e^{-\frac{E_l}{k_B T}}}{2} + \frac{e^{-\frac{E_u}{k_B T}}}{2}& 0 & 0 \\
0 & 0 & \frac{e^{-\frac{E_l}{k_B T}}}{2}+\frac{e^{-\frac{E_u}{k_B T}}}{2} & 0 \\
-\frac{e^{-\frac{E_l}{k_B T}}}{2}+\frac{e^{-\frac{E_u}{k_B T}}}{2} & 0 & 0 & 0 
\end{bmatrix}.
\end{eqnarray}
\end{widetext}

The four eigenvalues of the matrix $\rho^{\Gamma_{m},JC}$ can be determined and we find three strictly positive eigenvalues and only one negative,
\begin{eqnarray}
\lambda^-=\frac{1}{2}\bigg(e^{-\frac{E_g}{k_B T}}
-\sqrt{e^{-\frac{2E_g}{k_B T}}+\Big(e^{-\frac{E_l}{k_B T}}-e^{-\frac{E_u}{k_B T}}\Big)^2}\bigg),
\end{eqnarray}
which then enters the logarithmic negativity measure leading to Eq. (\ref{eq:negjc}).
\end{document}

\preprint{APS/123-QED}

%%%%%%%%%%%%%%%%%%%%%%%%%%%%%%%%%%%%%%%%%%%%%%%%%%%%%%%%%%%%%%%%%%
%                      Title and Authors                         %
%%%%%%%%%%%%%%%%%%%%%%%%%%%%%%%%%%%%%%%%%%%%%%%%%%%%%%%%%%%%%%%%%%

\title{SUPPORTING INFORMATION}
  \author{Dominik Sidler}
  \email{dsidler@mpsd.mpg.de}
  \affiliation{Max Planck Institute for the Structure and Dynamics of Matter and Center for Free-Electron Laser Science \& Department of Physics, Luruper Chaussee 149, 22761 Hamburg, Germany}
  \author{Michael Ruggenthaler}
  \email{michael.ruggenthaler@mpsd.mpg.de}
  \affiliation{Max Planck Institute for the Structure and Dynamics of Matter and Center for Free-Electron Laser Science \& Department of Physics, Luruper Chaussee 149, 22761 Hamburg, Germany}
  \author{Angel Rubio}
  \email{angel.rubio@mpsd.mpg.de}
  \affiliation{Max Planck Institute for the Structure and Dynamics of Matter and Center for Free-Electron Laser Science \& Department of Physics, Luruper Chaussee 149, 22761 Hamburg, Germany}
  \affiliation{Center for Computational Quantum Physics, Flatiron Institute, 162 5th Avenue, New York, NY 10010, USA}
  \affiliation{Nano-Bio Spectroscopy Group, Universidad del Pais Vasco, 20018 San Sebastian, Spain}

%\abbreviations{IR,NMR,UV}

%\begin{tocentry}
%\centering
%\includegraphics[width=1\linewidth]{figures/Abstract_fig_v2_red.pdf}
%\end{tocentry}

%\keywords{Cavity, Polariton, Dressed light-matter interaction, Pauli-Fierz Hamiltonian,  Exact Diagonalization, HD+, Temperature, (Grand)-Canonical Ensemble}

%%%%%%%%%%%%%%%%%%%%%%%%%%%%%%%%%%%%%%%%%%%%%%%%%%%%%%%%%%%%%%%%%%
%                            Abstract                            %
%%%%%%%%%%%%%%%%%%%%%%%%%%%%%%%%%%%%%%%%%%%%%%%%%%%%%%%%%%%%%%%%%%
%\begin{abstract}
%Goal: Focus on finite matter / photon subsystem temperature for $T=0$, when coupled to a cavity. 
%Show effect for Temperature $T>0$ and external pumping. Cooling / heating! Include COM to matter system. Letter Form!
%\end{abstract}

\date{\today}

\maketitle
 
%%%%%%%%%%%%%%%%%%%%%%%%%%%%%%%%%%%%%%%%%%%%%%%%%%%%%%%%%%%%%%%%%%
%                        Introduction                            %
%%%%%%%%%%%%%%%%%%%%%%%%%%%%%%%%%%%%%%%%%%%%%%%%%%%%%%%%%%%%%%%%%%
%\begin{widetext}
\section{Simulation Parameters}
For the numerical solution, parameters for HD$^+$ in a cavity were taken from Ref. \citenum{sidler2020chemistry} (e.g. particle mass or radial grid scaling) if not explicitly mentioned otherwise. The number  of radial grid points, was slightly reduced from $N_{\rm matter}=12$ to $N_{\rm matter}=10$, since we required additional resources to explicitly account for the ensemble averaging in $k_z$ (i.e. thermal effects acting on the charged COM motion).
The COM integrals were usually approximated by either $s=5$ (fluctuations) or $s=9$ (subsystem temperatures and entanglement) grid points, which effectively required either 3  or 5  evaluations of the computationally expensive $h(k_z)$ expression only. This simplification arises to the ensemble symmetry of $k_z$ with respect to the origin $k_z=0$, (i.e. $\mu=0$). The number of Fock states to represent the quantized field was chosen to be $N_{\rm pt}=4$ for the fluctuation analysis, $N_{\rm pt}=3$ for the subsystem temperature and $N_{\rm pt}=2$ for the entanglement measurements. For our chosen coupling strengths a choice $N_{\rm pt}>2$ only becomes relevant if one is interested in small deviations of the field fluctuations. Numerical convergence was ensured for all our results.

\begin{figure}
\begin{subfigure}[c]{0.48\textwidth}
\includegraphics[width=1\textwidth]{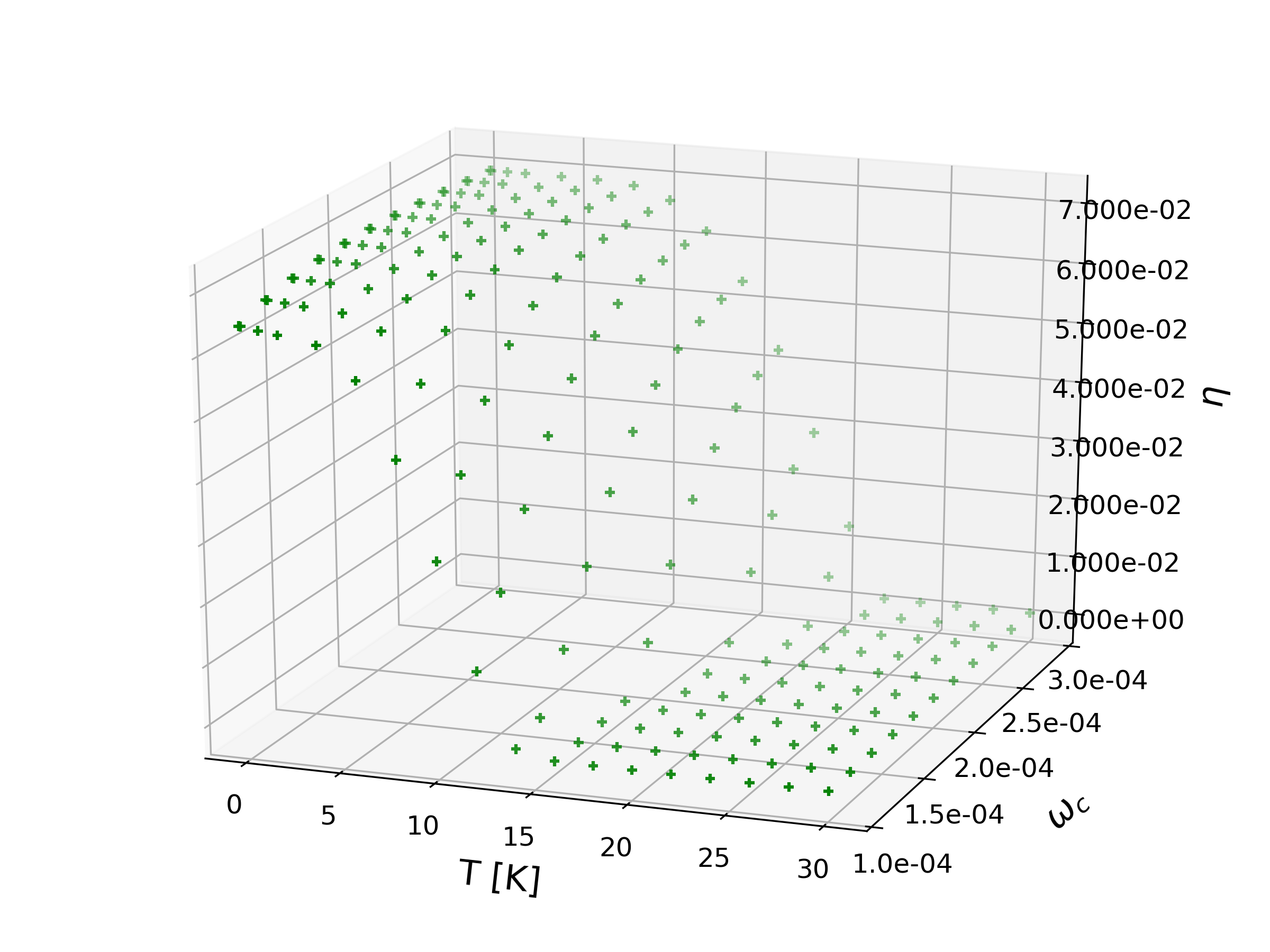}
%\subcaption{Subfigure Bild Nr. 1}
\end{subfigure}
\begin{subfigure}[c]{0.48\textwidth}
\includegraphics[width=1\textwidth]{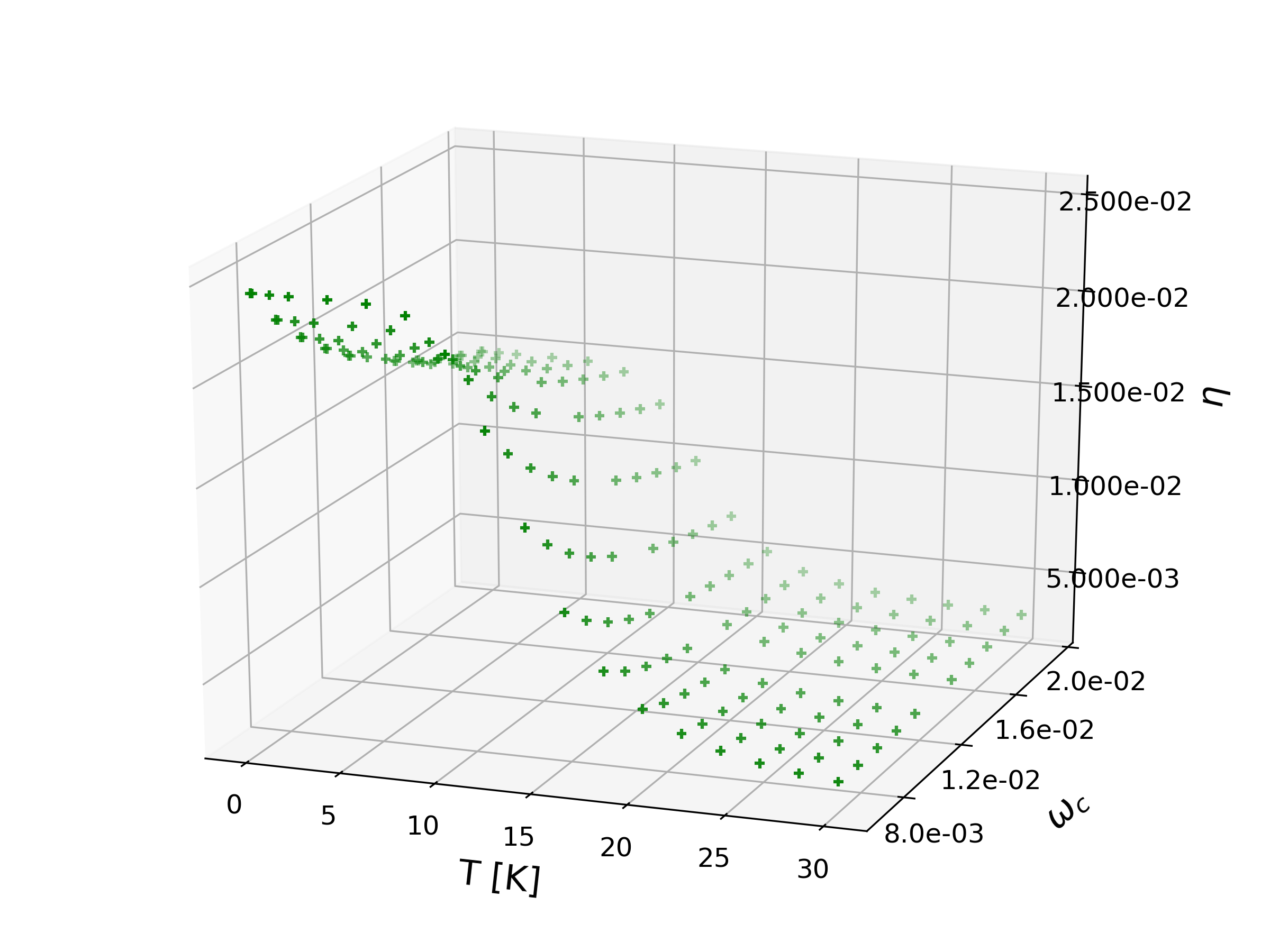}
%\subcaption{Subfigure Bild Nr. 1}
\end{subfigure}
\caption{Light-matter entanglement measured by logarithmic negativity $\eta$ with respect to the temperature $T$ for a coupling constant $\lambda=0.005$ at different cavity frequencies $\omega_c$ (given in atomic units). On the left, the cavity frequency is tuned close to the first ro-vibrational excitation, whereas on the right the IR to visible regime is covered. In both cases, light-matter entanglement is lost quickly in the deep cryogenic regime. However, while the numerical results for the first case are certainly converged, deviations may occur in the second case  due to the highly optimized grid-representation, which was designed to reproduce groundstate and ro-vibrational matter properties with high accuracy, but not simultaneously (!) the higher electronic and vibrational excitations. These states cannot be populated anyways in the chosen temperature regime, but still the accuracy of the matter basis contributions, which are mixed into the hybrid groundstate, may still be reduced in principle. Nevertheless, the loss of light-matter entanglement at low cryogenic temperatures seems to be a generic property for thermal ensembles under strong coupling conditions, roughly independently of the chosen cavity frequency. This is in line with typical experimental evidence, which does not allow for quantum  computing devices at sizable temperatures. }
\end{figure}

\bibliography{si}